\documentclass[12pt,preprint]{aastex}
\pdfoutput=1

\newcommand{\rfilter}{R$\rm _{filter}$}
\newcommand{\rcorrfilter}{r$\rm _{filter}$}
\newcommand{\rHeII}{R$\rm _{F469N}$}
\newcommand{\rhb}{R$\rm _{F487N}$}
\newcommand{\roiii}{R$\rm _{F502N}$}
\newcommand{\rcont}{R$\rm _{F547M}$}
\newcommand{\rha}{R$\rm _{F656N}$}
\newcommand{\rnii}{R$\rm _{F658N}$}
\newcommand{\rsii}{R$\rm _{F673N}$}

\newcommand{\roup}{R$\rm _{FQ437N}$}
\newcommand{\rniiup}{R$\rm _{F575N}$}
\newcommand{\rslow}{R$\rm _{FQ672N}$}

\newcommand{\Kfilter}{K$\rm _{filter}$}

\newcommand{\Ha}{H$\alpha$}

\newcommand{\Hb}{H$\beta$}
\newcommand{\Hg}{H$\gamma$}
\newcommand{\htwo}{H$_{2}$}

\newcommand{\sbunits}{photons~s$^{-1}$~cm$^{-2}$~sr$^{-1}$}

\newcommand{\kms}{km~s{$^{-1}$}}

\newcommand{\cmq}{cm$^{-3}$}

\newcommand{\pers}{s$^{-1}$}

\newcommand{\per}{$^{\rm{-1}}$}

\newcommand{\oiii}{[O~III]}
\newcommand{\oii}{[O~II]}

\newcommand{\nii}{[N~II]}
\newcommand{\sii}{[S~II]}

\newcommand{\Ne}{n$\rm_{e}$}
\newcommand{\Te}{T$\rm_{e}$}

\newcommand{\W}{\rm W_{filter}}
\newcommand{\Wc}{\rm W_{F547M}}

\newcommand{\Tc}{\rm T(max,F547M)}

\newcommand{\Tm}{\rm T(max,filter)}

\newcommand{\ratio}{\rm k}
\newcommand{\ngc}{the Ring Nebula}
\newcommand{\hezone}{He$^{++}$+H$^{+}$}
\newcommand{\ozone}{He$^{+}$+H$^{+}$}
\newcommand{\nzone}{He$\rm ^{o}$+H$^{+}$}
\newcommand{\Vexp}{V$\rm _{exp}$}

\slugcomment{To appear in the Astronomical Journal}

\shorttitle{NGC 6720 Imaging and Structure}
\shortauthors{O'Dell}

\begin{document}

\title{Studies of NGC 6720 with Calibrated HST WFC3 Emission-Line Filter Images--I: Structure and Evolution\
\footnote{
Based on observations with the NASA/ESA Hubble Space Telescope,
obtained at the Space Telescope Science Institute, which is operated by
the Association of Universities for Research in Astronomy, Inc., under
NASA Contract No. NAS 5-26555.}
\footnote{Based on observations at the San Pedro Martir Observatory operated by the  Universidad Nacional Aut\'onoma de M\'exico.}}

\author{C. R. O'Dell}
\affil{Department of Physics and Astronomy, Vanderbilt University, Box 1807-B, Nashville, TN 37235}

\author{G. J. Ferland}
\affil{Department of Physics and Astronomy, University of Kentucky, Lexington, KY 40506}

\author{W. J. Henney}
\affil{Centro de Radioastronom\'{\i}a y Astrof\'{\i}sica, Universidad Nacional Aut\'onoma de M\'exico, Apartado Postal 3-72,
58090 Morelia, Michaoac\'an, M\'exico}

\and

\author{M. Peimbert}
\affil{Instituto de Astronomia, Universidad Nacional Aut\'onoma de M\'exico, Apdo, Postal 70-264, 04510 M\'exico D. F., M\'exico}

\email{cr.odell@vanderbilt.edu}

\begin{abstract}

We have performed a detailed analysis of  \ngc\ (NGC~6720) using HST WFC3 images and derived a new 3-D model.  Existing high spectral resolution spectra played an important supplementary role in our modeling.  It is shown that the Main Ring of the nebula is an ionization-bounded irregular non- symmetric disk with a central cavity and perpendicular extended lobes pointed almost towards the observer. The faint outer halos are determined to be fossil radiation, i.e. radiation from gas ionized in an earlier stage of the nebula when it was not ionization bounded.

The narrow-band WFC3  filters that isolate some of the emission-lines  are affected by broadening on their short wavelength side and all the filters were calibrated using ground-based spectra. The filter calibration  results are presented in an appendix.

\end{abstract}

\section{Background and Introduction}
\label{sec:intro}

The Ring Nebula (NGC~6720, M~57) appears on the sky as an elliptical ring in most images. The ring form is due to the fact that He~II emission, which arises in the central regions is usually not imaged or is faint compared with the brightest optical lines. The bright, lower ionization region is usually called the Main Ring even though it must always be remembered that we are seeing a three dimensional object projected on the plane of the sky. The Main Ring is surrounded by a highly structured inner halo and a nearly circular outer halo of about 115\arcsec\ radius \citep{bal92,gue97}. The distance is uncertain and is best determined as 740$^{+400}_{-200}$ pc (O'Dell et al. (2009), henceforth OHS). The central star temperature is about 120000 K  (O'Dell et al. (2007b), henceforth OSH) and its luminosity is 200 L$_{\sun}$ (OHS). It has been argued (OSH)  that the nebula is old enough (7000 yrs since the end of the asymptotic giant branch wind) that it was previously fully ionized by the then more luminous central star, but has become ionization-bounded as the central star's ionizing luminosity has dropped, but it now may be in a stage when the ionization boundary continues to grow due to the decreasing density of the gas.  It is an appropriate object to address with the powerful WFC3 of the Hubble Space Telescope (HST). It's unique set of narrow-band emission-line filters are especially useful, but, they demand accurate calibration in the configuration within which they operate.

The three dimensional model that applies to the nebula has been the subject of numerous studies, with the most recent (O'Dell et al. (2007b, henceforth OSH) arguing that the nebula is a triaxial ellipsoid with a dense toroidal ring and low density axial caps, a refinement of the earlier models of Lame \&\ Pogge (1994) and  Guerrero et al. (1977, henceforth GMC). However, we show in \S\ \ref{sec:3D} that the best model is quite different those previously derived. 

In this paper (\S\ ~\ref{sec:NewObs} and appendix A) we present a detailed calibration of many of the WFC3 narrow-band filters using regions of \ngc\ that have been spectrophotometrically calibrated from ground-based observations. We then use this calibration to characterize the ionization structure (\S\ \ref{sec:IonStruct}) in the entire Main Ring,  elaborate on the detailed structure of the nebula (\S\ \ref{sec:details}), and derive a new and more accurate 3-D model (\S\ \ref{sec:New3D}). In the succeeding paper we derive local conditions of electron temperature and density and quantitatively analyze the small-scale temperature fluctuations. 

\section{New Observations}
\label{sec:NewObs}
The installation of the WFC3 camera on the Hubble Space Telescope (HST) during the final servicing mission in October, 2008 has provided a powerful new instrument for the HST user community. Although its field of view is less than that of the ACS camera, it possesses a wider variety of filters, and images a similar field of view as its predecessor axial instrument the WFPC2 with smaller pixels (0.0396\arcsec $\times$ 0.046\arcsec). The filter set includes narrow-band filters that isolate many of the brightest emission-lines of gaseous nebulae, including lines particularly useful for determining the physical conditions in planetary nebulae and H~II regions. 

\begin{figure}
\epsscale{0.75}
\plotone{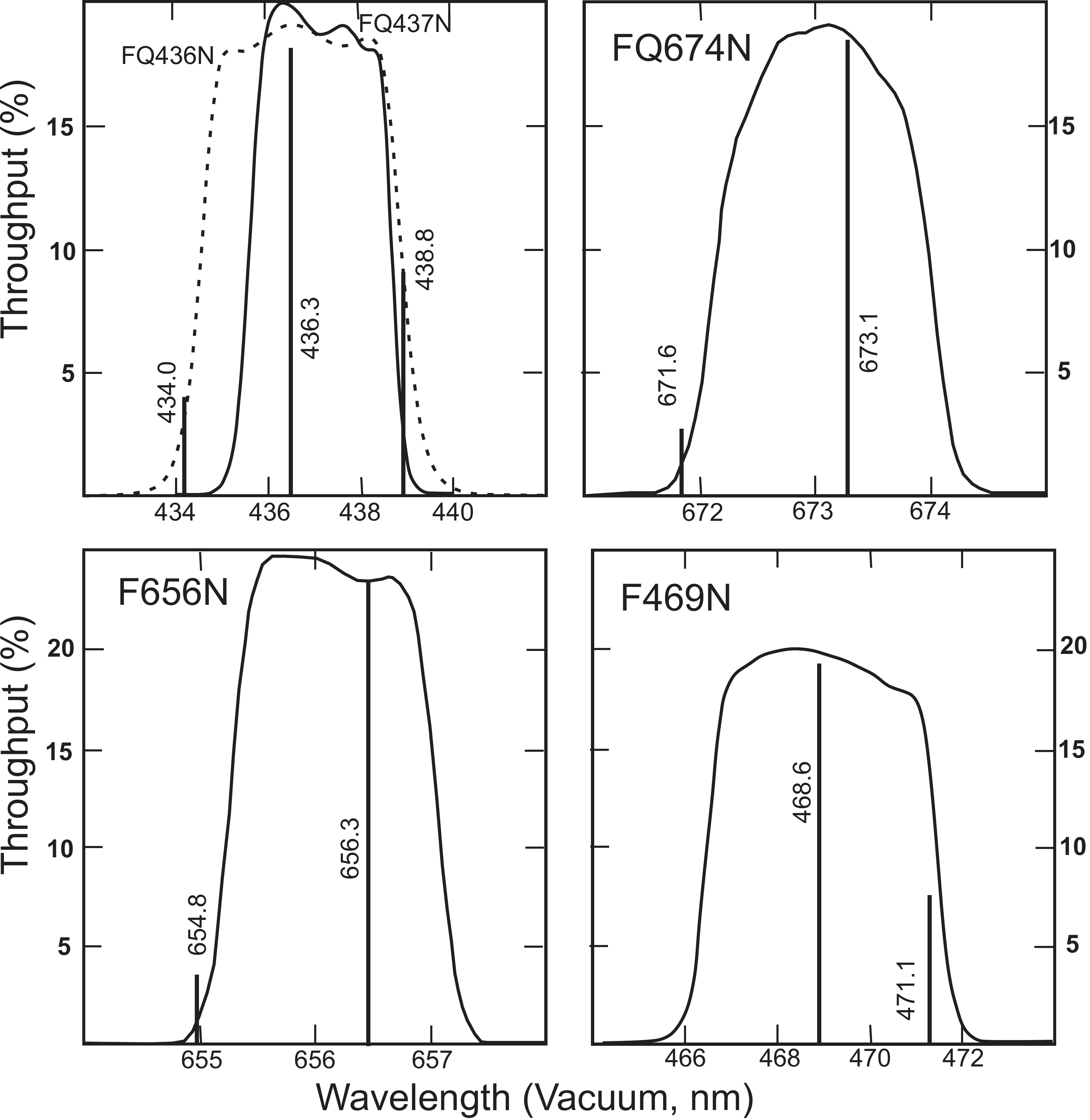}
\caption{These figures each plot the Throughput versus wavelength of the fived filters in our program that are affected by contaminating emission-lines. In each case we indicate the targeted emission-line and the contaminating line(s), without an indication of their true relative values. Vacuum wavelengths were used for plotting the emission-lines, but their designations are given in the more commonly used air-wavelengths. The Throughput values are from the WFC3 Instrument Handbook Appendix A and were determined from pre-launch tests. The short wavelength profile of the filter curves will have been altered due to the convergent beam of the WFC3.  The upper-left panel shows Throughput curves for the overlapping FQ436N (dashed line) and FQ437N(solid line) filters.
\label{fig:f1}}
\end{figure}

Although the throughput (T) of the HST-WFC3 has been determined from on-orbit observations in the wide-bandpass filters, the narrow-band filter calibration depends upon scaling ground-based transmission profiles using the wide-bandpass filter results. The narrow-band filters require an independent detailed calibration because the filter transmission profile is altered from that found in the ground-based determination. This is because the filters as-installed operate in a convergent beam, which slightly broadens the filter on the short wavelength side \citep{pbh98,lof11}. Although the targeted lines lie safely in the middle of the transmission profile of each filter, there are some cases where contaminating lines are also present (Figure~\ref{fig:f1}). In addition, emission-lines from gaseous nebulae are accompanied by a continuum and this produces a second source of a contamination. Use of these filters demands correction for both the contaminating lines and the underlying continuum. 

The methodology used for the calibration is similar to that employed previously in calibration of the WFPC2  \citep{ode09} and ACS \citep{ode04} emission-line filters. Because of the superior isolation of lines by the WFC3 filters, a simpler method of creating and applying the calibration was possible. The methodology and results are described in Appendix A.

The WFC3-UVIS (ultraviolet-visual) mode was designed with two types of narrow-band filters. What was expected to be the most frequently used filters encompass the full field of view set by the detectors (about 166\arcsec\ x 175\arcsec). In addition to these filters there are ensembles of four smaller filters put into an array that would otherwise be occupied by a full size filter. They are mounted with a central support between them, which obscures a central cross shaped portion. This means that for these "quadrant" filters, the unvignetted field of view is about 1/6th that of a non-quadrant filter. 

\begin{figure}
\epsscale{0.75}
\plotone{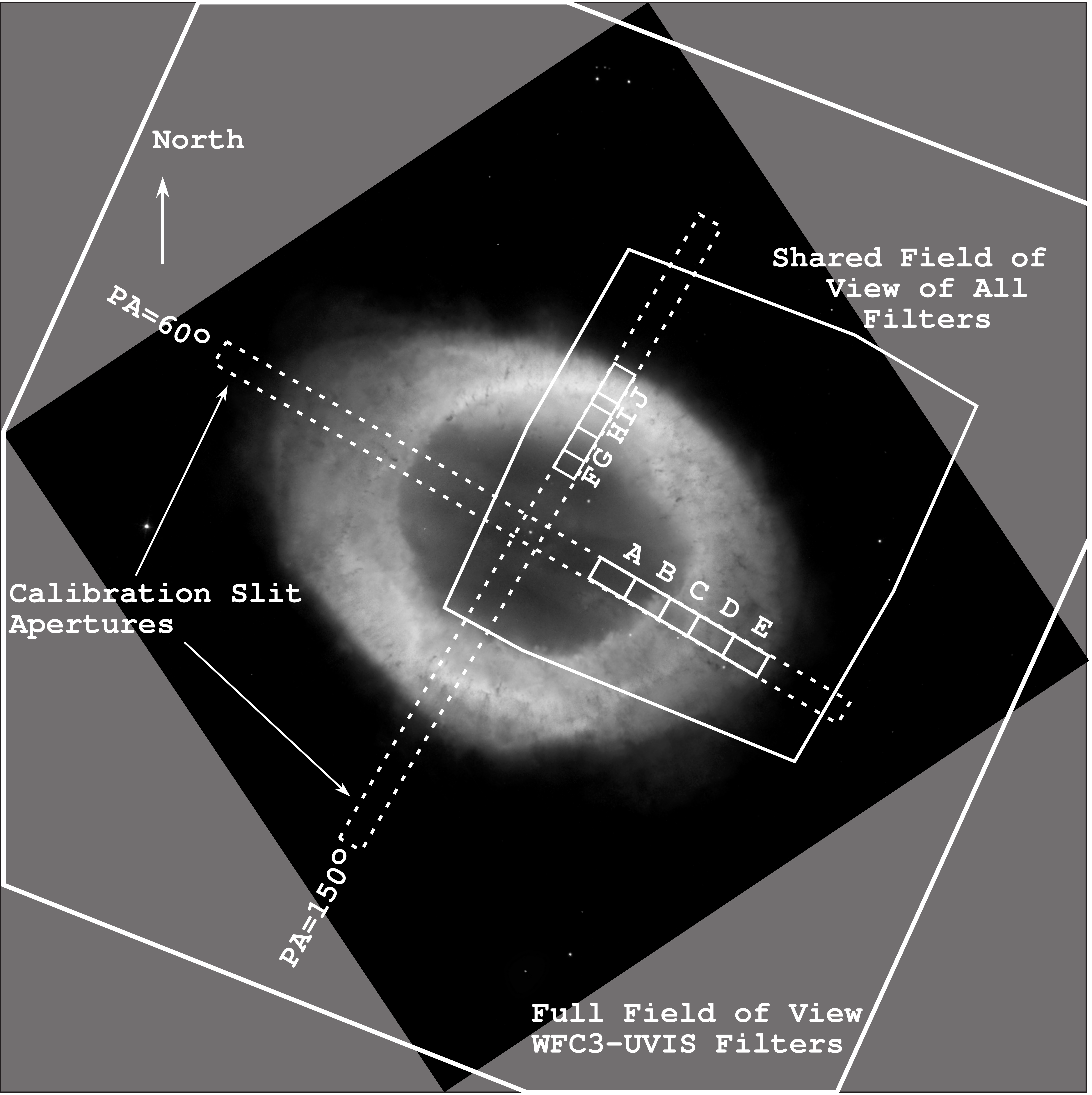}
\caption{This 170\arcsec\ x 170\arcsec\ image of the Ring Nebula has the vertical axis pointed towards north and is taken from the Hubble Heritage 1999-01news release of the STScI  and is a composite of HST WFPC2 images in the F469N (He~II), F502N (\oiii), and F658N (\nii) filters.  The outer solid lines show the limits of the WFC3-UVIS full-field filter images( F469N, F487N, F502N, F547M, F656N,  F658N, and F673N). The inner solid lines show the common fields of view of the quadrant filter images (FQ436N, FQ437N, FQ575N, FQ672N, and FQ673N). The dashed-line rectangles are the slit locations for the ground-based San Pedro Martir spectra and the small solid line rectangles show the regions used in calibration of the WFC3 images.
\label{fig:f2}}
\end{figure}

The non-quadrant filter images were centered on the Central Star of the planetary nebula and included the full bright inner ring of the nebula. The quadrant filter images were made at offsets to the northeast and the unvignetted images overlapped with one another and the non-quadrant filter images. This shared field of view is also shown in Figure~\ref{fig:f2}, with this area being 1572 x 1736 pixels of 0.04\arcsec\ or 63\arcsec\ x 69\arcsec. These HST program GO~12309  observations were made with an orientation so that the pipe-line processed images provided by the Space Telescope Science Institute had a positive Y axis pointed towards PA = 335\arcdeg. 
This orientation was determined by the availability of guide-stars and is close to the minor axis of \ngc\ as shown in Figure~\ref{fig:f2}. In a few  portions of this article (\S\ \ref{sec:IonStruct} and \S\ 2.4 of the succeeding paper)we report on samples along the orthogonal axes of the original image but usually report on analysis of portions of the images lying along and perpendicular to the minor axis, both for reasons of clarity of analysis and to agree with the ground-based slit observations. The first visit by the HST was on 2011 September 19 and the second on 2011 September 25. The first visit's filters, targeted emission lines and total exposure times were F469N (He~II 468.6 nm, 1400 s), F487N (\Hb\ 486.1~nm, 1400 s), F502N (\oiii\ 500.7~nm, 1230 s), F547M (a medium width filter with a central wavelength near 547.0~nm, 1230 s), F656N (\Ha\ 656.3~nm, 1120 s),
F658N (\nii\ 658.3~nm, 1400 s), F673N ([S~II] 671.6~nm + 673.1~nm, 1230 s), and FQ750N (continuum, 1400 s). The FQ750N filter image suffered from fringing and was not used in this study. The second visit was entirely of quadrant filter observations; FQ436N (\Hg\ 434.0~nm + \oiii\ 436.3~nm, 2600 s), FQ437N (\oiii\ 436.3~nm, 2640 s), FQ575N (\nii\ 575.5~nm, 2640 s), FQ672N (\sii\ 671.6~nm, 1270 s), and FQ674N (\sii\ 673.1~nm, 1270 s). Since the FQ436N filter was strongly contaminated by \Hg, it was not used in our analysis. Figure~\ref{fig:f2} also shows the locations of the ground-based slit spectra used for the calibration of the emission-line filters, see for reference \S\ \ref{sec:SPM}. 

\subsection{Ionization Structure}
\label{sec:IonStruct}

The ionization structure of the two-dimensional images of \ngc\ provides insight into the 3-D physical structure of the nebula. In this section we present the two-dimensional images and in \S\ \ref{sec:3D} discuss what this tells us about the 3-D structure of the object. 

\begin{figure}
\epsscale{1.0}
\plotone{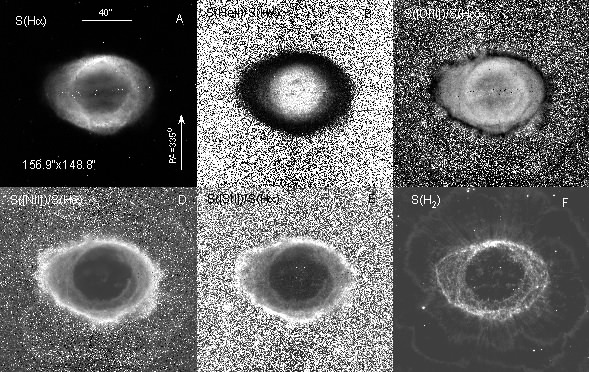}
\caption{This array of images is of a 156.9\arcsec x148.8\arcsec\ subsample from the full field of view narrow band emission-line filters and have been corrected for contaminating lines and continuum. The ``stitching'' across the middle is from where the gap between the two CCDs of the detector where filled-in by dithering the pointing. Panel A presents the  \Ha\ image  while the other panels depict ratios in various lines, except for panel F, which is a direct image in H$\rm _{2}$.  The H$\rm _{2}$ image was made with the Large Binocular Telescope and was provided by David Thompson. 
\label{fig:f3}}
\end{figure}

The expected sequence of ionization states for a central-star photoionized gas structure has been described multiple times, but for clarity of argument is represented here. Balmer recombination lines (including our \Ha\ and \Hb\ lines) are expected throughout the ionized gas. If there is a ionization front, the transition to neutral hydrogen occurs rapidly and at a well defined boundary called the ionization front. 
These two properties make \Ha\ emission a good reference source and is shown in Panel A of Figure~\ref{fig:f3}.  In the case of a hot Central Star, which is the case for \ngc\ with an effective temperature of about 120000 K (multiple results for the temperature are summarized in O'Dell et al. 2007), closest to the star there will be a He$^{++}$+H$^{+}$ zone. This zone uniquely produces He~II recombination lines, including the 468.6 nm line we observed. The ratio of He~II to \Ha\ is shown in panel B.  The apparent outer low positive values are due to small-scale errors in the zero points of the calibration and do not represent He~II emission. Further from the Central Star will be a zone of He$^{+}$+H$^{+}$ and the lines defining it are the \oiii\ nebular lines. In Panel C we show the ratio image for \oiii\ 500.7 nm over \Ha.  Further still from the Central Star will be a He$\rm ^{o}$+H$^{+}$ zone, best traced by the \nii\ nebular lines. In Panel D we show the ratio image for \nii\ 658.3 nm over \Ha. Because of the very high temperature of the central star, the \nii\ emitting zone is expected to be very narrow.  \sii\ emission occurs in a narrow region near and within the ionization front where there are some sulphur atoms that have not been ionized and there are sufficient energetic electrons from the ionization of hydrogen to excite them by collisions. The combined \sii\ nebular lines ratio to \Ha\ is shown in Panel E. The \nii\ and \sii\ emitting zones are expected to be narrow and are most visible when they are seen edge-on. Beyond the ionization front there will a photon dominated region (PDR) where molecules such as H$_{2}$ can exist and in Panel F we show a high resolution \htwo\ 2.12 \micron\ emission-line image made with the Large Binocular Telescope and provided by David Thompson.  

\begin{figure}
\epsscale{1.0}
\plotone{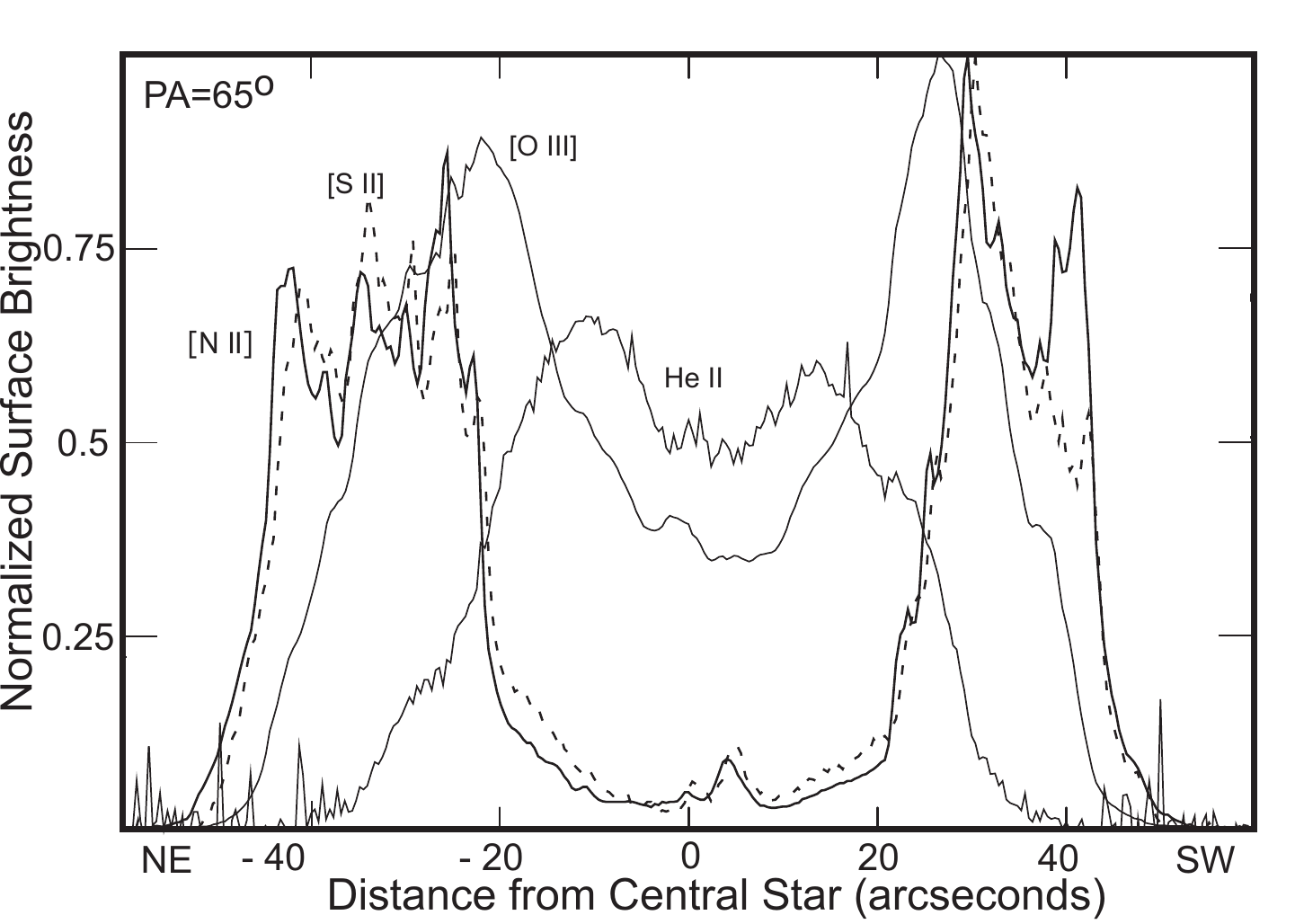}
\caption{These profiles along the semi-major axis sample of the full field of view filters depict the strongest emission-line of each ion. Again the samples were 4\arcsec\ wide and were averaged in 0.4\arcsec\ bins. The data are normalized to peak values near unity, except for He~II, which is normalized to a lower value for clarity.  The dashed line indicates the sum of the [S~II] 671.6 nm + 673.1 nm emission-lines.
\label{fig:f4}}
\end{figure}

\begin{figure}
\epsscale{1.0}
\plotone{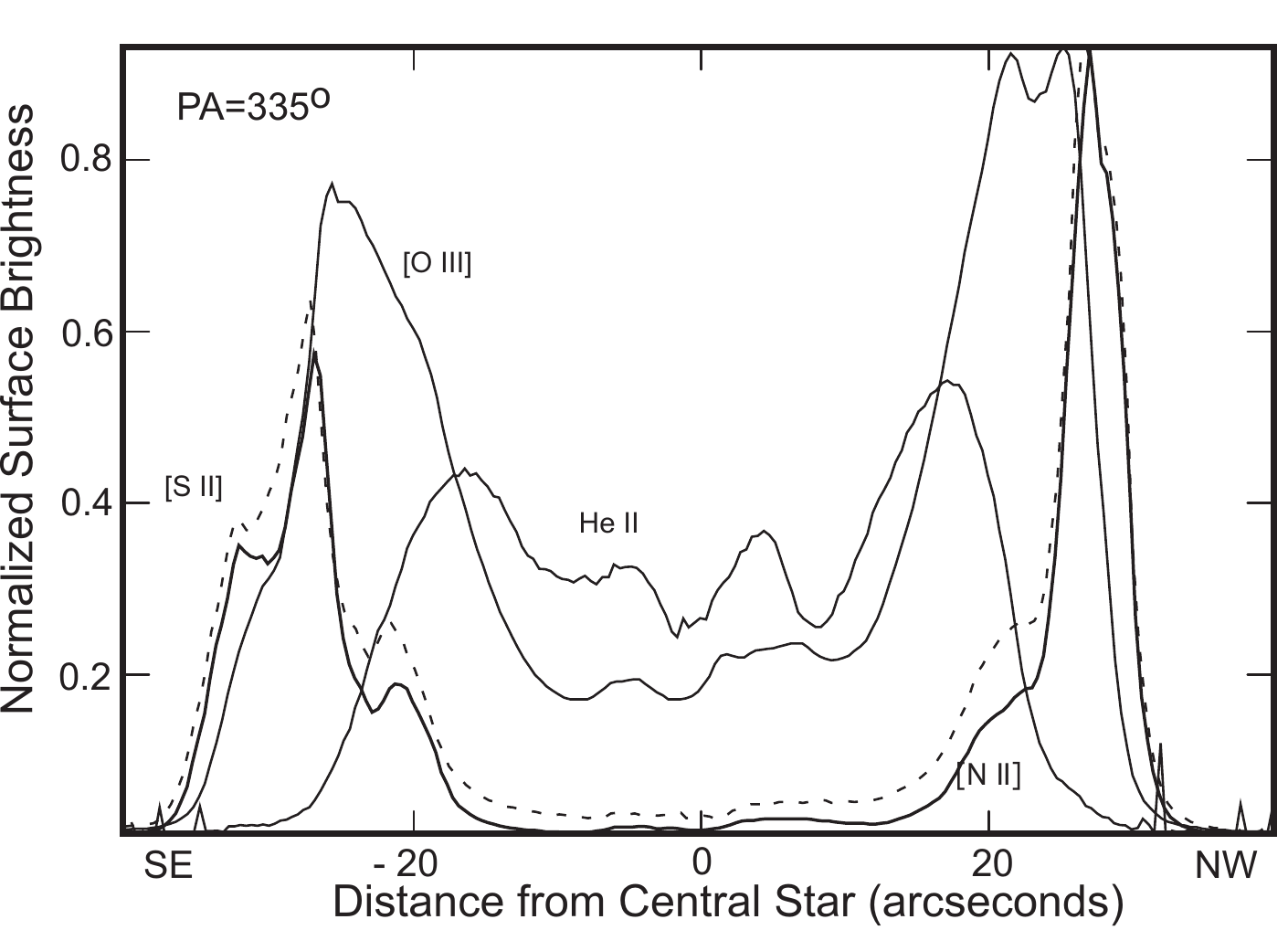}
\caption{This figure was prepared in the same manner as Figure~\ref{fig:f4} except it is for the semi-minor axis profiles. 
\label{fig:f5}}
\end{figure}

In Figure~\ref{fig:f4} and Figure~\ref{fig:f5} we present profiles in the observed line surface brightnesses at PA values of 65\arcdeg\ and 335\arcdeg\ from the images made with the full field of view (FOV) filters. These angles are close to the semi-major and semi-minor axes of \ngc. These profiles along the two key radial lines show in more detail what the images of Figure~\ref{fig:f3} demonstrate. There is a single central He~II emitting zone in the form of a central cavity surrounded by a ring of He~II emission that diminishes outwards. Outside of this one sees a rapid change in \oiii, \nii\ and \sii\ emission. The overly simplified interpretation is that the nebula is ionization bounded at about 42\arcsec\ along the semi-major axis and at about 30\arcsec\ along the semi-minor axis.  However, evidence for a more complex structure is clearly present. Along PA = 65\arcdeg\ we see that there are three clearly delineated ionization fronts seen in profile to the NE (at 44\arcsec , 36\arcsec , and 25\arcsec) and at least two (at 40\arcsec\ and 29\arcsec), with a possible third at 25\arcsec) to the SW.
Along PA = 335\arcdeg\ one sees evidence for ionization fronts seen in profile at three distances to the SE (33\arcsec , 28\arcsec , and 21\arcsec)  and two (at 21\arcsec and 27\arcsec) to the NW.

The projected \nii\ emitting zones are characteristically about 4\arcsec\ wide, which imposes an upper limit to the physical thickness of this boundary region.  The \sii\ emission occurs essentially along the same lines-of-sight (LOS) as the \nii\ emission, which could be explained as being due to projection effects when viewing a curved ionization front from side-on and the fact that the intrinsic width of the \nii\ emitting zone is expected to be very narrow because of the high temperature of the central star. We note that in the NE portion of the PA = 65\arcdeg\ profile we see that the peak of \nii\ emission occurs further from the Central Star than the \sii\ emission, a puzzling fact that must be related to the local geometry of the nebula  We also find that the \sii\ image of the nebula is more diffuse than that in \nii, which is contrary with the usual appearance of nebulae. This is   probably due to the lower signal to noise ratio of the \sii\ data and the fact that the intrinsic width of the \nzone\ zone must be quite narrow for this hot of a central star..

The presence of emission from mutually incompatible ions, e.g. He~II emission overlapping \nii\ emission, indicates that we are seeing in some regions that there are ionization fronts lying nearly perpendicular to the observer's LOS. This conclusion is strengthened upon examination of panel F of Figure~\ref{fig:f3}, where we see that the \htwo\ emission is not confined to the regions where we see ionization fronts nearly edge-on, but overlaps much of the optical line Main Ring. The evidence for seeing multiple ionization fronts at a variety of orientations is a key element in understanding the 3-D structure of \ngc. This is discussed in \S\ \ref{sec:3D}.

\subsection{Detailed Structures}
\label{sec:details}

\subsubsection{Knots}
\label{sec:Knots}

A survey of the nearest planetary nebulae with the HST \citep{ode02} indicated that small high-extinction features are probably always there and a natural part of the evolution of a planetary nebula shell. The cause of these features is unknown, with several instability mechanisms \citep{cap73,vis94} arguing for their formation within the expanding shell, probably at the ionization front. The dust component is probably there before the instability mechanism compresses the gas and with it the dust. A very different proposed origin \citep{dys89} is that these are condensations of material that existed in the extended atmosphere of the precursor star, although the absence of knots in very compact planetary nebulae argues against \citep{hug02b}. 
The features, subsequently called knots in this paper, have molecular cores \citep{hug02a}. They have been best studied in the Helix Nebula (NGC~7293). At a distance of about 219 $^{+27}_{-21}$pc \citep{har07}, the Helix is about one-third the  740$^{+400}_{-200}$ (OHS) distance to \ngc, offering a corresponding improvement in spatial resolution.  In the Helix Nebula the knots are seen as optically thick nearly circular dark cores with photoionization along the side facing the Central Star \citep{ohf07}. 
An ionization shadow is cast beyond the knot and the material being shed from the photoionized front side and streaming away from the Central Star is ionized in a sheath by scattered Lyman continuum photons, which gives the knots the appearance of having tails pointing away from the Central Star. One does not know from direct measurements the masses of the Helix knots, but under common assumptions about the properties of the dust particles, molecular fractions, and the gas/dust ratio, in addition to the count of the ubiquitous knots, it is likely that as much mass lies within the knots as in the photoionized gas (O'Dell et~al. 2007a and citations therein) . Whether these knots survive after injection of the nebular shell into the general interstellar medium is not known \citep{ob97}, but if they do, they would become an important component of the interstellar medium and the building blocks of new stars and proto-planetary disks.  For these reasons, it is important to study the knots in \ngc.

\begin{figure}
\epsscale{1.0}
\plotone{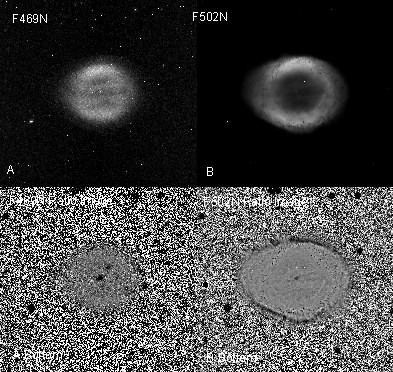}
\caption{The same FOV as Figure~\ref{fig:f3} is shown in each panel. Panels A and B are the
images in the F469N and F502N flters. Panels A Bottom and B Bottom are the upper
images divided by a gaussian blurred version of itself with a FWHM of 2.00\arcsec. The ratio images show best the fine scale structure.
\label{fig:f6}}
\end{figure}

\begin{figure}
\epsscale{1.0}
\plotone{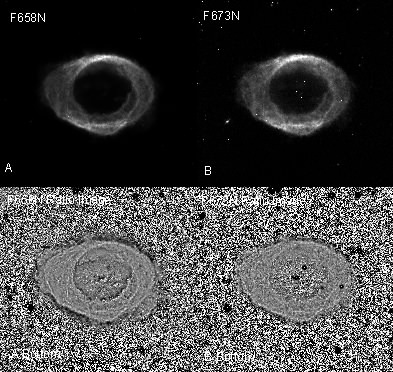}
\caption{Like Figure~\ref{fig:f6} except for the F658N and F673N images.
\label{fig:f7}}
\end{figure}

We noted the presence of knots in all of our WFC3 images, as shown in Panels A--E of Figure~\ref{fig:f3}. In order to enhance their visibility, we blurred the full FOV images with the IRAF\footnote{IRAF is distributed by the National Optical Astronomy Observatories, which is operated by the Association of Universities for Research in Astronomy, Inc.\ under cooperative agreement with the National Science foundation.} 
task ``gauss'' to a FWHM of 2.0\arcsec, then divided the original image by the blurred image. The resulting image has almost nulled-out the large-scale features of the nebula, rendering small features, both bright and dark, as more visible. The ratio images, over a range from 0.5--1.5, are shown in Panels C and D in Figure~\ref{fig:f6} and Figure~\ref{fig:f7}.  

Visibility of a knot  in silhouette (a dark knot) demands that the object lies closer to the observer than some of the gas producing the emission-line that has been isolated by the filter. A knot lying ahead of all of the emitting gas will have its maximum possible optical depth, whereas the same knot located near the far-side of the emitting column would have less apparent optical depth.  Of course a knot lying beyond all of the emitting gas would not be visible by this ``in-silhouette'' technique.  Even though \ngc\ must have a central concentration of material (argued to being essentially a bipolar disk seen nearly pole on by OSH, \citep{bal92, bry94} an alternative model of a ``bubbling prolate ellipsoidal shell'' has also been proposed in GMC). The presence of low ionization features at small angular distances from the Central Star is direct evidence of material lying well above and below the equatorial plane and the presence of H$_{2}$ emission in the Main Ring is difficult to explain. An equatorial concentration of material means that knots confined to one zone of ionization (as described in \S\ \ref{sec:IonStruct}) can appear outside their parent zone of ionization in projection on the sky. For example, even if knots were present only at the ionization front, they could be seen in projection against the higher ionization zones.  However, because of the known concentration of material towards the equatorial plane, there should be a concentration of objects towards the projected equatorial zones of their true location.  

A careful examination of the direct and ratio images in Figure~\ref{fig:f6} and Figure~\ref{fig:f7} in a sequence of decreasing ionization shows that one begins to see dark knots in the outer parts of the F469N images, in addition to one group beginning at PA = 218\arcdeg\ and $\theta$ = 13\arcsec\ (where $\theta$ is angular distance from the Central Star). There is then a big jump in the number of dark knots with the F502N images. In fact, there are no obvious cases where new dark knots are added in the F658N and F673N images, although the visibility of the furthest F502N objects improves in the lowest ionization images. We see a transition with ionization of the appearance of the dark knots. They are clearly dark in both the He~II and \oiii\ images whereas in general in \nii\ and \sii\ they appear bright, with this property varying between different arcs. 
 The knots commonly appear bright in \htwo\ emission, as seen in panel F of Figure~\ref{fig:f3}. This raises the possibility that they are excited by the same mechanism operating in the knots in the Helix Nebula \citep{hen07}.

The distribution of the dark knots are largely along arcs. These arcs are usually associated with a rapid drop in the F502N \oiii\ surface brightness, which must indicate the presence of a transition from a \hezone\  to a \ozone\ zone. These ionization transition arcs almost always have a nearby associated \ozone\  to \nzone\  transition zone and then indications of an ionization front. The concentration of knots to the arcs almost certainly indicates that the knots are product of an ionization front instability, rather than being broadly distributed within the stellar ejecta, as would be expected with an origin in the precursor star's atmosphere.   

At first examination the dark knots resemble those seen in the Helix Nebula in that both show a basic elongation along radial lines towards the Central Star. The clearly separated dark knots have an angular size of about 0.2\arcsec, corresponding to 150 AU, which is about one-half the linear size of the innermost Helix Nebula dark knots. They are probably the products of the same shaping mechanisms, that would have operated longer in the Helix Nebula (inner ring 6600 yrs, \citep{omm04}) than in \ngc\ (dynamic age 4300 yrs, OHS). Like the Helix Nebula there is not a bright rim on the star-facing side of the dark knots that would indicate their location within the \hezone. Unlike the Helix Nebula, there are not well-defined ionization rims in \Ha, \nii, and \sii\ emission with all the dark knots. When present they are about half (90 AU) the thickness of the Helix Nebula bright rims.  

When present, the ionized rims have thicknesses of about 0.1\arcsec\ (70 AU), only slightly larger than the WFC3 diffraction limit for sharp sources. These sizes are smaller than the 0.6\arcsec\ (130 AU) \citep{ob97} bright rims in the Helix Nebula. The lower visibility of bright rims with the \ngc\ knots is puzzling. The rims in \ngc\ are of lower surface brightness than those in the Helix Nebula, even though the photoionization process that produces both the basic surface brightness of the nebulae and the knots is the same and one would expect that the contrast against the nebula would be similar. The answer may lie in the marginal spatial resolution allowed at the distance of \ngc. The knots commonly appear bright in \htwo\ emission, as seen in panel F of Figure~\ref{fig:f3}. This raises the possibility that they are excited by the same mechanism operating in the knots in the Helix Nebula \citep{hen07}.

Although the \ngc\  features beyond the  knots show radial symmetry, they lack the well defined symmetry that characterizes the knots in the Helix Nebula. In a few cases it appears that the objects are irregular, the best example being the innermost dark knot at $\theta$ = 20.5\arcsec\ and PA = 266\arcdeg. However, it is most likely that what we are seeing is the nearly radial alignment of several independent knots.

\subsubsection{Inner Region Features}
\label{sec:inner}

\begin{figure}
\epsscale{0.5}
\plotone{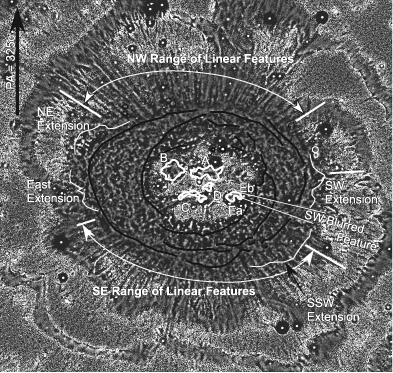}
\caption{Like the individual panels of Figure~\ref{fig:f6} and Figure~\ref{fig:f7} except starting with the \htwo\ panel of Figure~\ref{fig:f3}. The dark lines indicate the positions of arcs that lay along portions of the main ionization front seen in projection and the irregular light white lines the four regions of arcuate \nii\ emission that lie outside the boundaries of the Main Ring. Open Circles show the position of \nii\ bright knots that lie outside the boundaries of the Main Ring.  The range of positions over which linear features (Rays) are seen in the \oiii\ and \htwo\ images are indicated. The single Ray occurring outside these boundaries is indicated by a line and occurs just above the label for the ``SW Extension''. The position of the SW Blurred Feature is also shown. That feature and the Extension features are best seen in panel D of Figure~\ref{fig:f3} and bottom panels of Figure~\ref{fig:f7}. The irregular heavy white solid lines (A--Eb) outline the strong Inner Region low-ionization emission features.
\label{fig:f8}}
\end{figure}

An additional and problematic feature that is visible in the \Ha\ image in Figure~\ref{fig:f3} and the He~II and \nii] images in Figure~\ref{fig:f6} is the presence of bright broad streaks of emission that about lie about 5\arcsec\ NE and SW of the central star and are almost parallel to the major axis of the Main Ring and superimposed on the central drop in surface brightness.  These features are broad in He~II, then show progressively more structure with lower stages of ionization. In \nii\ they resemble the inner region features labeled A through Eb in the \htwo\ image (Figure~\ref{fig:f8}. The low-ionization appearance of these inner features led GMC to call them ``Bubbles''.  The full range of conditions (molecular hydrogen through He~II) indicates that they are neutral features surrounded by photoionized gas. The nature of these features is discussed in \S\ \ref{sec:InnerFeatures}. 

\subsubsection{The SW Blurred Feature}

In Figure~\ref{fig:f7} we see clearly a feature partially visible in Figure~\ref{fig:f3} and Figure~\ref{fig:f8} and designated as the ``SW Blurred Feature''. It is a divergent width region of quite diffuse emission bounded by the lines shown in Figure~\ref{fig:f8}. It appears to originate in the group of features lying 16\arcsec\ towards PA = 214\arcdeg\ from the central star, designated as object Eb in Figure~\ref{fig:f8}. The region of origin appears as a sub-group of dark knots with emission-line bright edges facing the central star, with a similar sub-group a few seconds of arc to its SE (object Ea).  Where it crosses the Main Ring it appears to obscure the emission from the object and must lie in the foreground. It extends at least 16\arcsec\ beyond the outermost ionization boundary of the Main Ring. 

The radial velocity of the \nii\ emission-lines from the parent group of knots was determined from the average of slit spectra made at PA values of 110\arcdeg\ and 
120\arcdeg\ (c. f. \S\ \ref{sec:New3D}) to be -34.4 \kms\ with respect to the systemic velocity of 0 \kms\ LSR. This fortuitously convenient designation of velocities will be used throughout the remainder of this paper. This negative radial velocity indicates that the blurred feature lies on the near side of the Main Ring. This position argues that material in the SW Blurred Feature lies in the foreground and emission from it distorts the appearance of the Main Ring where the two components are seen. This interpretation does require that there is material lying outside the Main Ring and that this material is affected by radiation shadowing by the Eb inner feature.

\subsubsection{Features Outside the Main Ring}
\label{sec:outer}

\begin{figure}
\epsscale{1.0}
\plotone{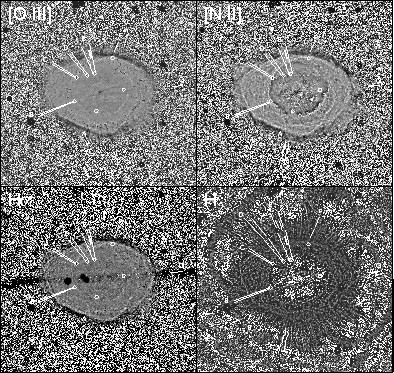}
\caption{Like the bottom panel of Figure~\ref{fig:f6}, Figure~\ref{fig:f7} A,  and Figure~\ref{fig:f8}, with the addition of a similarly processed \Ha\ image. The features are designated A-G in counterclockwise notation with the PA values being A(142.6\arcdeg ), B(65.8\arcdeg ), C(34.2\arcdeg ), D (13.2\arcdeg ), E(352.5\arcdeg ), F(315.1\arcdeg ) and G(248.8\arcdeg).  The solid lines outline the boundaries of the feature and the open circles indicate the associated dark knots. Narrow straight dashed lines indicate enclosed narrow dark features seen in the F502N images and the cross-hatched region within object E indicates an extended region dark in the F502N images. The heavy dashed line indicates the position of the \oiii\ portion of feature A. In both the \nii\ and \htwo\ images the wavy light line in feature A indicates the associated bright \htwo\ emission and in the \nii\ image the light solid line indicates the \nii\ emission. Full resolution images are given in the electronic version of this paper.
\label{fig:f9}}
\end{figure}

Examination of the monochromatic line ratio (Figure~\ref{fig:f3}) and the gaussian deblurred ratio images (Figure~\ref{fig:f6} and Figure~\ref{fig:f7}) indicates the presence of many radial ``Rays'' at the outer perimeter of the main ring. Similar Rays  have been seen in several other planetary nebulae, including IC~418 \citep{rl12} and NGC~6543 \citep{bal04}. The most similar object is the Helix Nebula \citep{ode00,omm04}. The explanation first advanced for the Helix Nebula and similar linear features in the Orion Nebula \citep{ode00} was that these are ionization shadows cast onto nebular gas. Within such a shadow gas would not be photoionized by radiation directly coming from the Central Star. This gas would only see ionizing radiation emitted by recombining ions, usually called the diffuse radiation field. The continuum of the recombining ions is much more concentrated towards the ionization edge than the stellar continuum. This means that the gas within the shadowed region is expected to have a lower electron temperature than the ambient gas. 

Initially the theory for shadows in planetary nebulae and within H~II regions was quite straightforward \citep{can98} and was later expanded \citep{wm04} for the case of illumination of the Diffuse Ionized Gas. O'Dell et al. (2004) pointed out that in the case of high ionization nebulae the situation was more complex because multiple stages of helium ionization could be present. If the shadowed region is optically thin to LyC radiation, then there will be emission from throughout the shadowed region, but it will be primarily from recombining H$^{+}$ and [O~II] and \nii. If the cylindrical shadowed region is optically thick to LyC radiation, an ionization front will be formed, inside of which \htwo\ emission may be observed and outside of which one will see hydrogen recombination lines such as \Ha and  forbidden \oii\ and \nii\ emission. This balance of emission-lines will extend out until direct illumination by the central star is reached. \oiii\ emission is not expected because where the diffuse emission includes the recombining He$^{++}$ continuum, it will move oxygen into the triply ionized state.  Recombining He$^{+}$ produces photons just above 24.6 eV and these are insufficient to doubly ionize oxygen (O'Dell et al. 2004, \S\ 4.2). 

In the case of the Helix Nebula the shadows immediately beyond the dark knots have been well studied \citep{ode00,ohf05,ohf07}. The shadows are optically thick to  LyC radiation at the edge of the shadow, so there is a sheath of ionized gas seen in \Ha\ and within in it a layer of \htwo\ 2.12 \micron\ emission.  The unexpected brightness of the arc facing the ionizing star was quantified in O'Dell et al. 2007 and explained in Henney et al. (2007).  As in the case of our study of \ngc, \oii\ was not observed, but \nii\ was and it was found that no \nii\ was observed, the only \nii\ radiation being accounted for by nebular emission scattered by dust in the shadow column. There was no \oiii\ emission. The shadowed regions beyond the dark knots in \ngc\ are much more irregular than in the case of the Helix Nebula and we have not attempted a similar analysis of them. In the case of the outer parts of the Helix Nebula there seems to be a background of \oiii\ emission punctuated by over 100 dark radial lines. 

We see similar outer structure in \ngc\ as shown in panel B-bottom of Figure~\ref{fig:f6}, although the Rays are not as numerous. Like the Helix Nebula, there is a fainter outer glow of extended \oiii\ emission against which shadows from the inner dark knots are seen. We have conducted a thorough examination of the sharpened optical emission-lines and the \htwo\ emission, showing the results in Figure~\ref{fig:f9}. One immediately sees that the Rays are best seen in \oiii\ and \htwo.  We have identified six outer features (Rays or combinations of Rays) that have a clear association with an inner knot or combination of knots. 
Since all of the knots are bright in \nii\ emission and objects on the further side of the nebula will appear only in emission (c.f. \S\ ~\ref{sec:New3D}), we had to search for a correlation between the outer Rays and knots and objects that would appear as dark knots if appropriately located. 
The Ring Nebula features have a very different structure than the well-studied shadowed regions within the Helix Nebula. In \ngc\ the common form is that within the projected shadow there is an ionized sheath (seen as nearly radial lines on both sides) in \nii\ and a nearly superimposed fainter and more diffuse feature in \Ha. Inside this ionized sheath is a layer of \htwo\ emission.  Except for feature A, there is no evidence of \oiii\ emission.  This appearance argues that in the outer parts of the nebula the shadowed regions are optically thick to the diffuse radiation field and that \htwo\ is present, either having recently formed or surviving from an earlier, denser phase of the nebula. Unfortunately, one cannot tell from the emission-lines present the nature of the surrounding material.

Feature A originates with a well defined dark knot whose apex is  16.0 \arcsec\ from the central star. This knot is not seen in \oiii\ and is faintly detectable in \Ha, but is bright in \nii\ and \htwo. There is a significant separation between the well-defined shadow-cone behind the knot and its appearance outside the Main Ring. On the counterclockwise edge one clearly sees \oiii\ emission, with a progression to slightly larger PA's with \nii, \Ha, and \htwo\ emission. On the larger PA edge one finds \htwo\ strong and only faint \nii\ is detected. The presence of \oiii\ emission on one side indicates that this feature is in a portion of the nebula illuminated by the stellar continuum, rather than by diffuse radiation. Because of the radial alignment, the ionizing radiation must be coming from light scattered by a nearby dusty region.

Feature F originates with a dark knot 28.6\arcsec\ from the central star. This dark knot is about 0.5\arcsec\ wide perpendicular to a radial line and is unusual in that it appears dark in \oiii, \nii, \Ha, and is barely detected in emission in \htwo.  This visibility places it at or on the observer's side of the main ionization front of the Main Ring. There is a narrow (about 0.1\arcsec) straight dark lane in \oiii\ traceable out to 9.9\arcsec\ from the dark knot. Beyond 5.9\arcsec\ from the dark knot one begins to see emission in \nii, \Ha, and \htwo, with this emission displaced about 0.3\arcsec counterclockwise. The absence of \oiii\ emission and the one-sided nature of the observed emission indicates that this shadowed gas is being illuminated by diffuse radiation from one side only.

Feature G is similar to Feature F in that it is a narrow straight feature that lies along a projection of an unresolved \htwo\ feature at 18.7\arcsec\ from the central star. This associated knot is unique in that it is the only source that has no ionized gas presence, indicating that it lies in a neutral region on the far side of the Main Ring. The noisy region where the CCD detectors merge preclude a detailed study of Feature G in either \Ha\ or \oiii. However, it is clearly present in \nii, with a width of about 0.1\arcsec. The unresolved \htwo\ emission starts at 53.5\arcsec\ from the central star and extends for about 12.1\arcsec.
 
All of these linear Rays exist or extend into regions outside of the Main Ring. The presence of some dark Rays in \oiii\ indicates that there is ionized material beyond the Main Ring, in contradiction with the fact that this material lies beyond the ionization front of the Main Ring. The explanation for this is discussed in \S\ \ref{sec:Halo}.

It is noteworthy and discussed in \S\ \ref{sec:Halo} that these linear Rays are found only over a limited range of PA values, not being found near the projections of the major axis of the Main Ring. The one clear exception to the exclusion rule is shown as a solid line above the SW Extension label in Figure~\ref{fig:f8} and is Feature G in Figure~\ref{fig:f9}.

There are also arcuate features seen outside the outer ionization boundary of the Main Ring. These are most clearly seen in the contrast enhanced images (Figure~\ref{fig:f6} and Figure~\ref{fig:f7}.  They are labeled ``Extensions'' (with direction indications) in Figure~\ref{fig:f8} and three of the four (SSW Extension is the exception)fall within in the major axis zone that are almost entirely devoid of Rays. They are higher ionization than the regions containing the rays, as measured by the \oiii/\nii\ ratio. They seem to be disrupted material in the region of the outer \oiii\ glow that surrounds the nebula (c. f. \S\ \ref{sec:Halo}).

\section{The 3-D Model of \ngc 's Main Ring}
\label{sec:3D}

A satisfactory model of the Ring Nebula must satisfy both its elliptical appearance on the plane of the sky and its velocity structure. Because rotational symmetry in the initial ejection of material from the Central Star and its acceleration is expected, the first models assumed that the ionized gas was an ellipsoid form. In the case of an oblate ellipsoid  (an ellipse rotated about its minor axis) the short axis of the observed ellipse would be due to foreshortening by the rotational axis being tipped with respect to the LOS. In the case of a prolate ellipsoid (an ellipse rotated about its major axis) the maximum observed ellipticity would occur when the rotational axis lies in the plane of the sky (perpendicular to the LOS) and the observed ellipticity would reduce to zero when the rotational axis is pointed along the LOS. Discrimination between the two forms is provided by radial velocity data along the apparent major and minor axes.  In the case of an intrinsic prolate ellipsoid the radial velocities along the major axis would differ on the two ends producing a tilted long-slit spectrum, with the red-shifted end in the direction of rotational axis pointed away from the observer. In the case of an oblate ellipsoid, the tilted spectrum would lie along the apparent minor axis, with the red-shifted end in the direction of the rotational axis that is pointed away from the observer.

The modern era of attempts to model \ngc\ can be traced to the study of Masson(1990) that established the basic ionization structure of \ngc . This could be explained if \ngc\ is a prolate spheroid with a concentration of material around its equatorial radius (this lies in the plane perpendicular to the rotational axis) and the rotational axis is tilted about 30\arcdeg\ with respect to the LOS.
This model was not tested with velocity data. The first subsequent study employing radial velocities \citep{bry94} argued that the dense equatorial ring was actually part of a bi-polar structure, where the faint polar regions 
flare out before becoming closed, whereas Masson's prolate ellipsoid monotonically closed).

Since Masson's paper there have been a succession of studies, most including some radial velocity data. The primary emission-line used was the \nii\ 658.3 nm line because the emitting layer is thin (hence the effects of any velocity gradients with distance are minimal) and the line suffers relatively little thermal broadening (5.7 \kms\ FWHM for \Te = 10000 K).  Multiple studies GMC, OSH\citep{bry94,ste07} have used the \nii\ line, all with long-slit spectroscopy, usually with a few slits centered on the Central Star and a few additional positions. La Palma observations \citep{bry94} were made at several positions (including one on the Central Star), including some observations also in the \oiii\ 500.7 nm and 656.3 nm \Ha\ lines. 
Spectroscopy from Calar Alto Observatory GMC employed three centered and two outer region settings. San Pedro M\'artir Observatory observations have been made \citep{ste07} with 10 slit positions, across the Main Ring, but only two settings included the Central Star.  

In terms of understanding the structure of the Main Ring, it is most important to determine the radial velocity distribution along multiple PA's and multiple emission lines. This allows determination of the spatial variations in both radial velocities and ionization structure. This is what lay behind the planning of a second San Pedro M\'artir study (OSH), where the entire Main Ring was observed with the \nii\ 658.3 nm, \Ha\ 656.3 nm,  and \oiii\ 500.7 nm lines in PA increments of 10\arcdeg, the He~I 468.6 nm line with 30\arcdeg\ increments, and the \sii\ doublet lines were observed only along the major and minor axis. 

It should be noted that small differences in the assumed PA for the long axis of \ngc\ will be found in the literature. Not only does the range reflect the judgement of the authors, but also the fact that the Main Ring is not symmetric on its NE and SW sides. We see in Figure~\ref{fig:f3} that to the SW it is more rounded and the orientation would be PA = 65\arcdeg , whereas the NE boundary is more pointed, with the greatest extension pointed towards PA = 55\arcdeg.  

\subsection{The most detailed earlier model of \ngc 's Main Ring.}
\label{sec:detailedPrevious3D}

In OSH a 3-D model of \ngc\ was created using the methodology developed at Padua University. In this approach flux calibrated long-slit spectra mapping the nebula in the plane of the sky are turned into a 3-D model under the assumption that all motion is radially outward and homologous. Because the Main Ring lies close to the plane of the sky, the scale of the expansion rate (\kms /arcsec) had to be determined under an assumption of the size of the nebula along the LOS. Within the additional assumption, which had reasonable constraints, that the ratio of the longest apparent axis to the nearly LOS axis was 1.5, they found the expansion relation to be \Vexp\ = 0.65 x R\arcsec , where \Vexp\ is the velocity of expansion (\kms) and R\arcsec\ the distance from the Central Star to the volume element expressed in equivalent angular scale.  

The resulting model confirmed several features of previous models, but made important improvements. The nebula must be a triaxial spheroid , rather than the biaxial structure inherent to a prolate or oblate ellipsoid. The longest apparent axis (which we will call the semi-major axis) is 44\arcsec\ [0.16 pc at 740 pc distance (OHS)] and lies along a line pointed to about PA = 60\arcdeg , the second axis (which we will call the semi-minor axis) is 30\arcsec\ (0.11 pc) and is perpendicular to the semi-major axis. These axes define the equatorial plane. The third axis is perpendicular to these axes and is about 59\arcsec\ (0.21 pc) and we will call this the semi-polar-axis. The corresponding full-length values of these axes we call the major, minor, and polar axes.
Noting that the radial velocity of the NE and SW tips of the \nii\ spectra along the major axis was 6 $\pm$ 2 \kms\ different (with the SW
end of the Main Ring moving away from the observer), OSH found that the polar-axis was tipped only 6.5\arcdeg\ $\pm$ 2\arcdeg\ from the LOS.  The much smaller intrinsic size of the semi-minor-axis is the simplest explanation, because if it were due to foreshortening due to tilt there would be a much larger velocity difference at the ends of the minor-axis spectra.  They also determined, like in earlier studies, that the density of gas was concentrated towards the equatorial plane and that the nebula is ionization bounded there. However, they concluded that because \nii\ emission was weak compared with \Ha\ emission near the Central Star that the nebula is optically thin to Lyman continuum radiation (LyC) along the polar axis.  In addition to material being concentrated towards the equatorial plane, it is also unevenly distributed along that plane, being greatest in the direction of the minor-axis,  and this accounts for the much higher surface brightness of the Main Ring on its NW and SE sides.  

\subsection{Observational results from infra-red and radio studies}
\label{sec:IRandRadio}

Although most studies of \ngc\ have been made by imaging and spectroscopy in optical wavelengths, there have been important contributions to the question of the 3-D structure from longer wavelength investigations, a single study in the radio window and multiple infra-red images in continuum, emission-line, and molecular emission. 

The single radio window investigation is that of Bachiller et al. (1989), who mapped \ngc\ in the CO molecule at 13\arcsec\ and 1.3 \kms\ resolution. This emission can only arise from a region shielded from ionizing and photo-dissociating radiation (i.e. optically thick to LyC and near ultraviolet radiation) and it is significant that they found that CO emission traces the optical Main Ring. This result, combined with our current knowledge of the presence of multiple neutral gas and dust cores \citep{ode02}, immediately argues that this molecular emission arises in the knots, c.f. \S~\ref{sec:Knots}.

The Ring Nebula has been imaged many times in the 2.12 \micron \htwo\ emission-line with increasing spatial resolution \citep{kas94, spe03,hir04} with the highest resolution images being those presented in Figure~\ref{fig:f3}. The Hiriart (2004) study has sufficient velocity resolution (24 \kms) to provide kinematic information.  Imaging in the dust continuum and lower excitation \htwo\ lines \citep{hora09,vanh10} has also been done. Most recently \citep{sah12} the SOFIA airborne observatory has be used to observe the nebula in the [C~II] 158 \micron\ line with eight usable samples of 15.6\arcsec\ and 2.9 \kms\ resolution. With ionizing photons of 11.3 eV necessary to remove the first electron from carbon and 24.4 eV to remove the second, one would expect the [C~II] emission to arise from within the PDR and extend through the \nii\ emitting region. 

\subsection{Derivation of a new model for the Main Ring using both imaging and spectroscopy.}
\label{sec:New3D}

We draw on our new WFC3 images and the spectroscopic results in OSH in deriving a new model for \ngc. In the present study we reassess the spectroscopic results and utilize them in a different fashion from the OSH paper.  We also draw on the results of the GMC study but different results for the correct model.

\begin{figure}
\epsscale{1.0}
\plotone{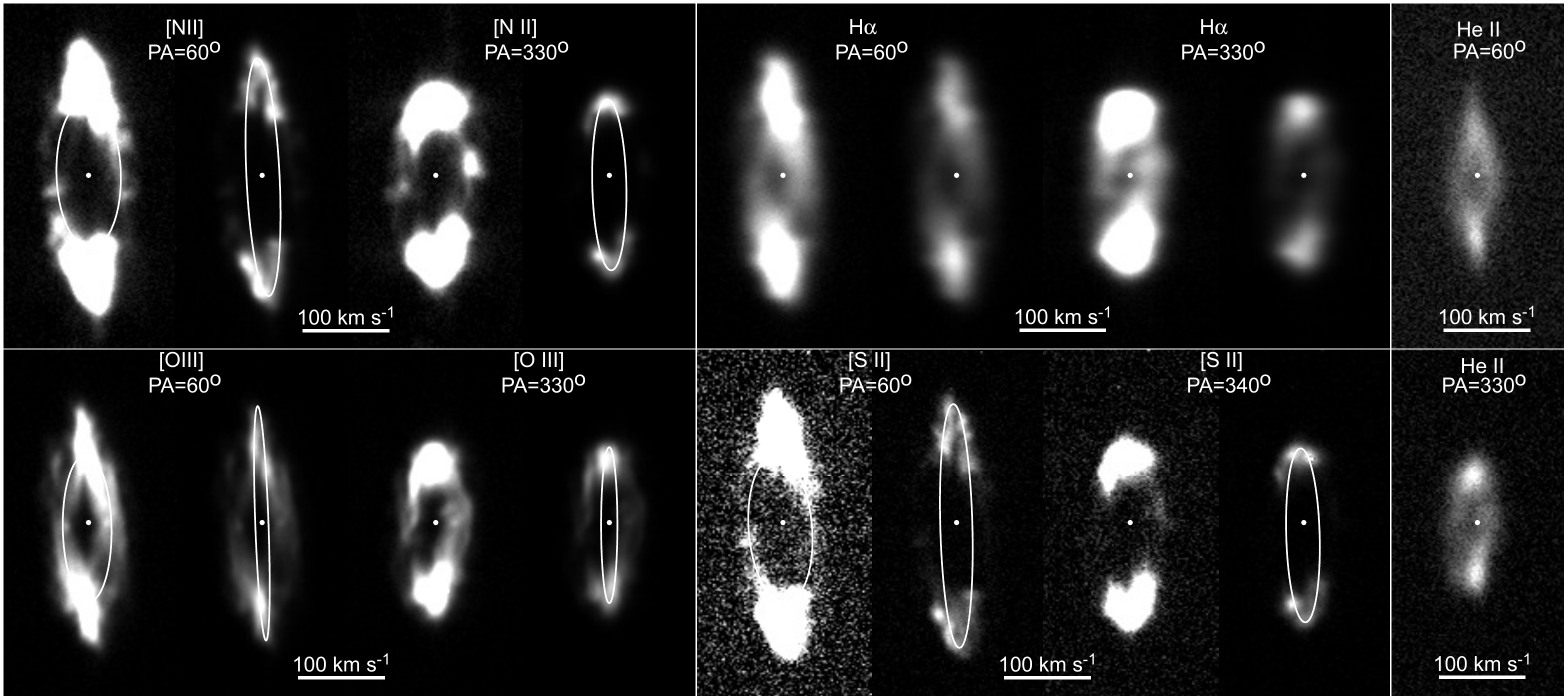}
\caption{Spectra averaged over 30\arcdeg\ intervals centered on the approximate semi-major (PA = 60\arcdeg) and semi-minor axes of the Main Ring are shown for the strongest emission-lines studied, drawing on the spectra from San Pedro M\'artir Observatory presented in OSH. Except for He~II, the spectra are presented at two levels, left to show the faintest features and the right the brightest features. Each spectrum shows 202 \kms\ along the X axis and 125\arcsec\ along the Y axis.
\label{fig:f10}}
\end{figure}

Because of the regular large-scale form of \ngc, we have averaged, when possible, the spectra over the 30\arcdeg\ intervals centered on the assumed semi-major (PA = 60\arcdeg) and  semi-minor (PA = 330\arcdeg) axes. These are presented Figure~\ref{fig:f10}. This figure shows all of the lines (except He~II 468.6 nm) in two levels of display, the right hand panels showing the brightest emission from the Main Ring and the left panels show the same spectrum with the Main Ring saturated but the dark central region spectra adequately displayed.  The \nii\ spectra are very similar to those of GMC, although our angular resolution is worse. Figure~\ref{fig:f10} demonstrates the much greater volume of data available than in previous studies because of the inclusion of multiple emission-lines covering the full range of ionization states and at positions radiating from the Central Star. These spectra show the usual trends of the ionized zone radii becoming larger with decreasing ionization energy and the velocity splitting also becoming larger with decreasing ionization energy. 

The presence of outer Main Ring emission in \sii\ and \nii\ makes it clear that the Main Ring is optically thick to LyC radiation and the shape of the line splitting indicates that the emission comes from successive zones of ionization, there being material of varying stages of ionization from near the Central Star right out to the ionization boundary.  The conclusion of a large optical depth in the LyC is reinforced by examination of Figure~\ref{fig:f3}, where we see that there is an \htwo\ emitting boundary around the Main Ring. However, if the Main Ring was simply a smooth ionized volume within a neutral circumstellar disk one would expect the brightness of the loops of emission at the top and bottom of the spectra to vary smoothly, since the surface brightness at the face of the ionization front would vary with the LyC flux from the Central Star. Narrow arcs of emission are not seen in \Ha\ since that radiation arises from throughout the ionized zone, but in all of the lower ionization energy ions we see that the narrow arcs are highly structured. 

Two alternative explanations for this irregularity appear possible. The first is that the gaps in the emission represent physical gaps in material and where LyC photons can escape. The second is that the brighter regions represent where the ionization front is tilted more nearly along the LOS. The second  explanation is preferred because one sees arcs of \htwo\ emission at or just beyond the ionization fronts identified in \ref{sec:IonStruct} and the \nii /\Ha\ ratio does not drop dramatically between the identified ionization fronts. The former indicates that the PDR is quite thin and is also seen in projection when the local portion of the ionization front is tipped towards or away from the observer while the latter is contrary to the expectation that gaps in the ionization front would drop significantly in surface brightness and the gas that is there would be quite high ionization. 

The usual method of interpreting spectra such as those in Figure~\ref{fig:f10} is that both the inner material and the loops  arising from the Main Ring are part of the same velocity system. We argue that this is not the case. We show in Figure~\ref{fig:f10} by superimposed ellipses how the data are better fit by different ellipsoids. A lower LOS \Vexp\ is appropriate to the Main Ring emission and a higher LOS \Vexp\ is appropriate to the central region emission, with the central region velocity-ellipse beginning just outside of the inner boundaries of the Main Ring.  The ellipses fitting the Main Ring spectra indicate a central region \Vexp\ = 19.0$\pm$2 \kms\ for the average of the \nii\ and \sii\ emission and 9$\pm$2  \kms\ for \oiii, while the ellipses fitting the central region emission give  \Vexp\ values of 37 \kms\ for \sii, 36 \kms\ for \nii, 28 \kms\ for \oiii, 27 \kms\ for \Ha, and 20 \kms\ for He~II. We assign this central emission region to the polar Lobes introduced in \S\ \ref{sec:New3D} and adopt an expansion velocity of 34 \kms\ for them. Clearly the central region emission and the Main Ring emission represent different velocity systems and the central region values are refined later in this section by taking averages over all PA's that were observed.  

It has been previously identified that the SW tip of the major-axis spectra are slightly redshifted. We determine the radial velocity difference of the NE-SW tips to be 13 \kms\ for \nii, and 11 \kms\ for \oiii. The \nii\ value is larger than the 9 \kms\ reported by Guererro (1997) and 6 \kms\ in OSH. The differences are probably due to the method of fitting the Main Ring spectra.  Previous studies found no tilts in the spectra along the minor-axis. This sense of tilt and the smaller minor-axis has led to the determination that the Main Ring emission is fit by a prolate ellipsoid with the SW tip pointed away from the observer.  GMC used their velocity tilt and arguments on the shape of the Main Ring to conclude that the major axis of the prolate ellipsoid is pointed 30\arcdeg\ away from the plane of the sky. OSH argues that the tilt is only 6.5\arcdeg$\pm$2\arcdeg\ and that the size of the minor-axis is determined by the object being triaxial.  At this point a discussion using ellipsoids is essentially moot because of the ambiguity of the forms. It the tilt is small, then the full major axis laying nearly in the plane of the sky is only slightly larger than 88\arcsec\ (0.32 pc) in length and the full minor axis nearly in the plane of the sky is 60\arcsec\ (0.22 pc).  If we use the velocity difference of 13 \kms, then our value of the tilt is 13.5\arcdeg\ and the physical length of the full major axis is 0.31 pc. We do see evidence for a small velocity difference along the minor-axis, with an average of 5$\pm$2 \kms. This indicates that the minor axis is tilted slightly in the SE direction. Given the small tilts of the major and minor axes, we will henceforth refer to the axis passing through the central region as the polar axis.

\begin{figure}
\epsscale{1.0}
\plotone{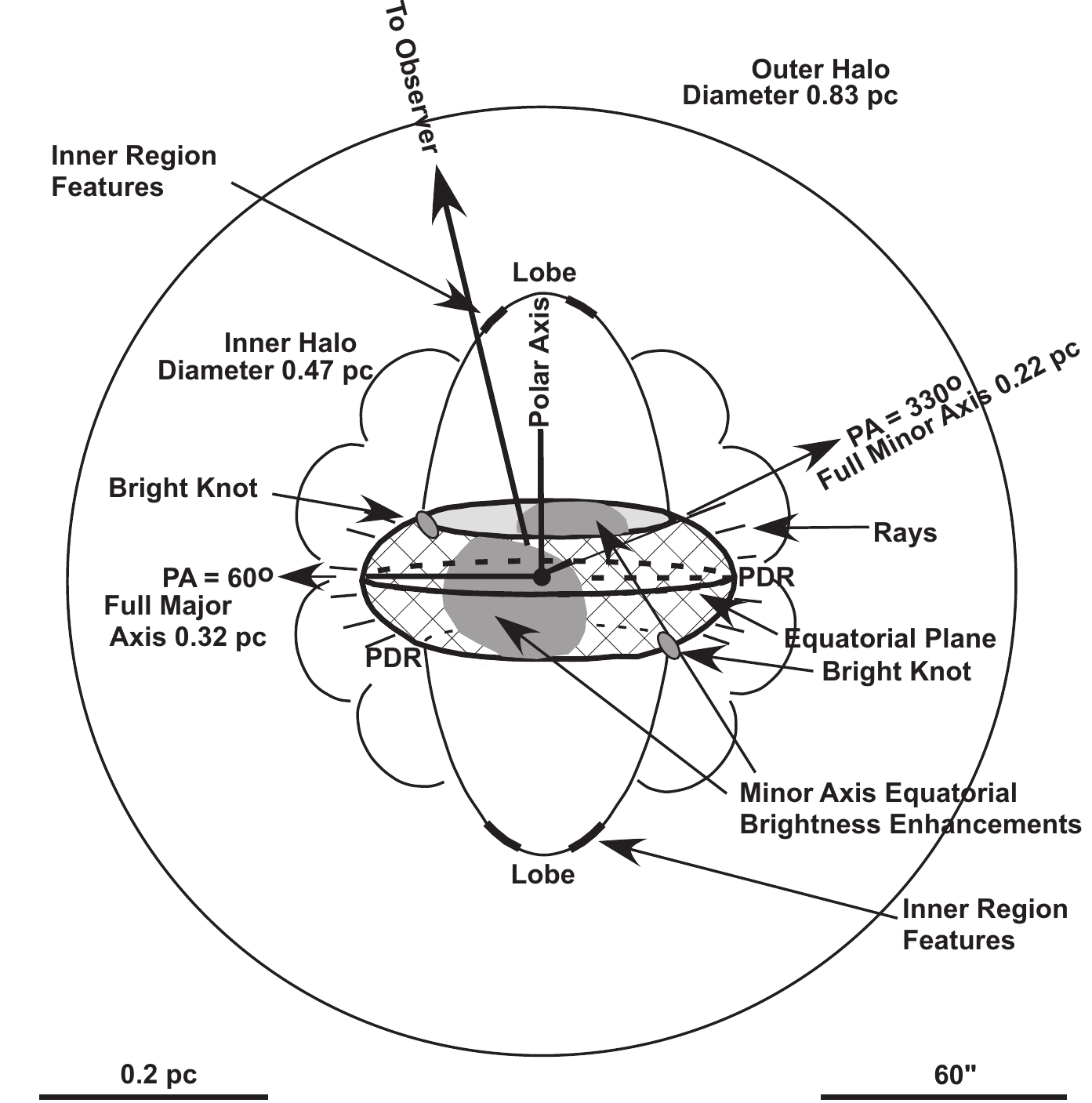}
\caption{A simplified depiction of our 3-D model for the Main Ring of \ngc\ is shown here. 
\label{fig:f11}}
\end{figure}

Our model for the 3-D structure of the Main Ring is shown in Figure~\ref{fig:f11}. It demonstrates the orientation and dimensions. The cross-hatched surface represents the ionization boundary within the equatorial plane and the flared features at the opening of the Main Ring represents where Lobe features along the polar axis begin. In no case of the OHS or GMC spectrophotometric samples do the intensities of emission-lines (\sii\ and [O~I]) arising from the transition  from the \hezone\ to the PDR drop to zero, even in the samples closest to the Central Star. This indicates that not only is the Main Ring ionization bounded, but this is also true for the Lobes.  The Bright Knot regions are where this model predicts that the low-ionization knots would appear because of project effects (one is looking essentially along the main ionization front. The location and polarity of the velocities agree with those seen in Figure~\ref{fig:f10}. 

The smaller size of the nebula in the equatorial plane along the minor axis (0.11 pc radius as compared with 0.16 pc radius along the major axis) accounts for the higher surface brightness of the Main Ring in the areas along the minor axis. As first detailed in Baldwin et al. (1991), a photoionized layer in front of an ionization front will have a surface brightness proportional to the flux of  LyC photons arriving at that front. This means that the surface brightness ratio in a Balmer line such as \Ha\ along the two axes should be about  (0.16/0.11)$^{2}$~=~2.1. Since the observed ratios are actually 1.9, this means that the difference in the size of the major and minor axes adequately accounts for the asymmetry in the surface brightness at different position angles around the Main Ring.

The visibility of the dark knots gives us further information about the geometry. Recall from \S\ \ref{sec:Knots} that they are always bright in \sii,  \nii, and \htwo, being most evident in the higher signal-to-noise ratio \nii\ images. This means that one can detect the presence of objects like those we call dark knots even when they fall on the far side of the nebula and there is no material to form a background against which a silhouette is seen.  We note in Figure~\ref{fig:f6}-Panel-B-bottom and Figure~\ref{fig:f7}-Panel-A-bottom that there are fainter arcs of knots on the inner parts of the semi-major and semi-minor axes with corresponding dark knots only faintly visible in the NW semi-minor axis arc. Since the dark knots are most visible when seen in projection against background material, we conclude that the opening that merges into the Lobe nearer the observer is wider than the opening on the far side. This asymmetry joins the multiple others for this object (e. g. unequal density distribution around the equator of the Main Ring and the asymmetry of the Main Ring).

As noted in \S\ \ref{sec:IRandRadio} the CO molecular emission \citep{bac89} peaks within the optical Main Ring image, indicating an association with the dusty knots that lie in its outer parts. In the velocity resolved \htwo\ study \citep{hir04} one sees a velocity distribution along a plane perpendicular to the sky and passing through PA = 33\arcdeg\ that the velocity splitting beyond distances of 30\arcsec\ mimic in shape and are slightly smaller in magnitude than the optical \nii\ line splitting. This is what is expected if the \htwo\ emission there is coming from the knots. The HST proper motion study of \ngc\ OHS concluded that the knots were moving with the gas. Beyond about 30\arcsec\ the \htwo\ line splitting increased to about 85 \kms, which is somewhat larger than the expansion velocities we derive in the Lobes. We interpret Hiriart's inner \htwo\ emission as arising from the PDR on the outside of the Lobe. Because of the lack of spatial resolution in the CO study and the low signal to noise ratio of the velocity resolved \htwo\ study, we won't refer further to their conclusions about models for the nebula in our comparison of various published models. The low spatial resolution (15.6\arcsec ) study of [C~II] \citep{sah12} is also not useful except to note that their maximum line splitting of 35 \kms\ is consistent with an origin in the outer parts of the Main Ring in the model presented here. 

\subsection{The Structure of \ngc\ Outside of the Main Ring}
\label{sec:Halo}
There are four components of \ngc\ beyond the Main Ring, the polar axis Lobes,  the region of \oiii\ glow on the outer boundary of the Main Ring, the Inner Halo, and the Outer Halo. The Lobes are difficult to see directly because they are seen almost along the polar axis. We have no evidence about the form of the Lobes. They could be bubbles as shown in Figure~\ref{fig:f11} or almost open, as seen in the well defined bipolar nebula M2-9 \citep{doy00}. Balick et al. (1992) also demonstrate that in the case of the bipolar nebula NGC 650-1 that the Lobes can be well defined bubbles and intermediate cases \citep{ode02} like IC~4406 also exist. As in the case of the Lobes, the components of the Inner and Outer Haloes are seen best where their ionization boundaries are viewed along the LOS. In fact, the Outer Halo (radius of 115\arcsec) is only seen as a circular ring with gaps, suggesting that it is a thin shell. The Inner Halo is composed of many arcuate structures (loops), suggesting that they are thin "bubbles". They are seen starting near the edge of the Main Ring, with almost all of them within a distance of 65\arcsec\ from the Central star. The fact that both the Inner Halo loops and the Outer Halo ring are visible in \htwo\ emission establishes that they are ionization bounded features.  

\begin{figure}
\epsscale{0.25}
\plotone{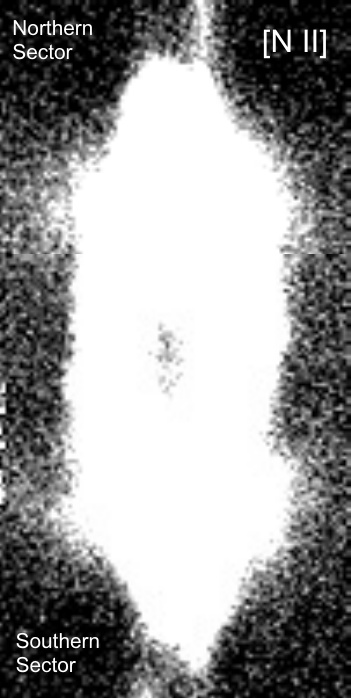}
\caption{
Averaged spectra over the entire nebula are shown for \nii, with the level of illumination set to show the features in the outer regions. The Northern Sector includes the samples between 240\arcdeg\ through north to 70\arcdeg. The Southern Sector includes the samples between 80\arcdeg\ through 230\arcdeg.
\label{fig:f12}}
\end{figure}

The radial velocities of the Halo features are difficult to determine because of their low surface brightnesses.  The only high resolution spectra are those of GMC that report long slit spectra passing through the Central Star and extending beyond the Outer Halo in addition to two slits displaced specifically to observed the Halos. Gurerrero et al. (1997) report that their \nii\ spectra show a broad component that narrows with increasing distance from the Central Star and there are additional narrower smaller components along its red and blue edges. They report the spread in velocities of the smaller and narrower components to be 25 \kms\ - 30 \kms.
In the region of transition from the Inner Halo to the Outer Halo they see the smaller components on both sides of the broad component.  Our examination of their Figure~8 is that the extended broad and narrowing component is actually a velocity ellipse with a splitting corresponding to an expansion velocity of about 15 \kms. This must be associated with the Outer Halo. The spread of velocities of the smaller features associated with the Inner Halo indicate that they are parts of an inner shell with a velocity splitting of 25 \kms - 30 \kms\ at their characteristic distances of 60\arcsec -80\arcsec. We present in Figure~\ref{fig:f12} our all PA's averaged \nii\ spectrum adjusted to show features beyond the Main Ring. There were no outer region features seen in the \sii, \Ha, \oiii, and He~II spectra. The \nii\ averaged spectrum indicates for the southern sector (PA's 80\arcdeg - 230\arcdeg) velocities at -17 \kms\ and +12 \kms\ at a distance from the Central Star of 61\arcsec. The northern sector (PA's 240\arcdeg through North to 70\arcdeg) give velocities at -11 \kms\ and +15 \kms\ at 58\arcsec. Our results are fully consistent with those of GMC and together to two data sets establish that the Inner Halo features correspond to a highly irregular shell of smaller features 
with a radial velocity difference of 28$\pm$2 \kms\ at Central Star distances of 60\arcsec -80\arcsec. This indicates that the expansion velocities of the Inner Halo and Outer Halo are very similar and are about 15 \kms.

Assuming that no deceleration has occurred, \Vexp (Halos) = 15 \kms\ correspond to ages of the 14380 yrs (Inner Halo) and 25442 yrs (Outer Halo). The age for the Main Ring is reported (OSH) as 7000 yrs. If the \nii\ expansion velocity for the central region (34 \kms) applies to the major axis, then the Main Ring's age would be 4314 yrs; however, it is likely that the expansion in the equatorial plane is slower than along the polar axis and this age is a lower limit.

Since both the Main Ring and the Lobes are optically thick to LyC radiation, the Inner Halo and Outer Halo features are no longer subject to photoionization. This means that the gas there is slowly recombining and that emission will die-out with the same timescale as the characteristic recombination timescale of 120000/\Ne\ yrs \citep{ost06}. OSH argue that \ngc\ was previously fully ionized when the Central Star was much more luminous in LyC photons. This would have also allowed the formation of ionization fronts in the outer features of \ngc. As the LyC luminosity dropped and the ionization front in the Main Ring formed LyC photons would no longer penetrate beyond the Main Ring. The Lobes would have become ionization bounded later than the Main Ring because of their lower densities, thus allowing the irradiation of the outer regions to be sustained longer. However, no photoionization must be occurring now, since both the Main Ring and Lobes are optically thick, and the ionization is what remains following the earlier phase in the evolution of the planetary nebula. 

Can the emission outside the Main Ring be fossil-radiation, that is radiation from material no longer being photoionized? The relative surface brightness of the Main Ring and a few limb-brightened Inner Halo arcs in \Ha\ were measured and was found to be 362. If the emissivities and path-lengths are the same, this corresponds to a density ratio of 19.0 and if a characteristic \Ne\ for the Main Ring is 500 \cmq, then the Inner Halo density is about 26.3 \cmq\ (without a path-length correction). The Inner Halo arcs look like thin shells seen in projection, therefore a path-length correction is needed and will be quite dependent on the geometry. If the path-length correction scales as the first power of the distance (the Main Ring is about 24\arcsec\ radius and the samples of the Inner Halo were about 84\arcsec), then the density is 14.0 and the decay timescale is 8570 yrs. If the path-length correction scales as the square of the distance, then the density is 7.5 \cmq\ and the decay timescale 16000 yrs.  Even the shorter of these timescales is long enough that the radiation we are seeing from the Inner Halo (and presumably the Outer Halo too, since it is even lower surface brightness and density) arcs are from gas that is no longer in photoionization equilibrium.  Recall that the OSH age since the end of the AGB stellar wind end was 7000 yrs.

Fossil ionization may also be the explanation of why we are able to see radiation shadows (the Rays)  in \oiii\ and there are the higher ionization extension features around the apparent rim of the Main Ring even though that object is obviously ionization-bounded today. These shadows would have been present even when the equatorial ring of material was optically thin. Now the Central Star has dropped dramatically in LyC luminosity, an ionization front has formed, leaving the low density gas outside the equatorial concentration to slowly recombine leaving within it the legacy of the shadowing while actively being photoionized. In this scenario the \oiii\ glow around the Main Ring represents the residual of the initial higher ionization region and the Inner Halo and Outer Halo are the residual of the initial low ionization regions. If the Rays are indeed fossil-radiation shadows cast by the knots, then this means that the Extensions are features beyond the main ionization front that somehow disturb the conditions that have allowed the radiation shadows to continue to the present.

\subsection{Features projected near the Central Star}
\label{sec:InnerFeatures}
\begin{figure}
\epsscale{0.6}
\plotone{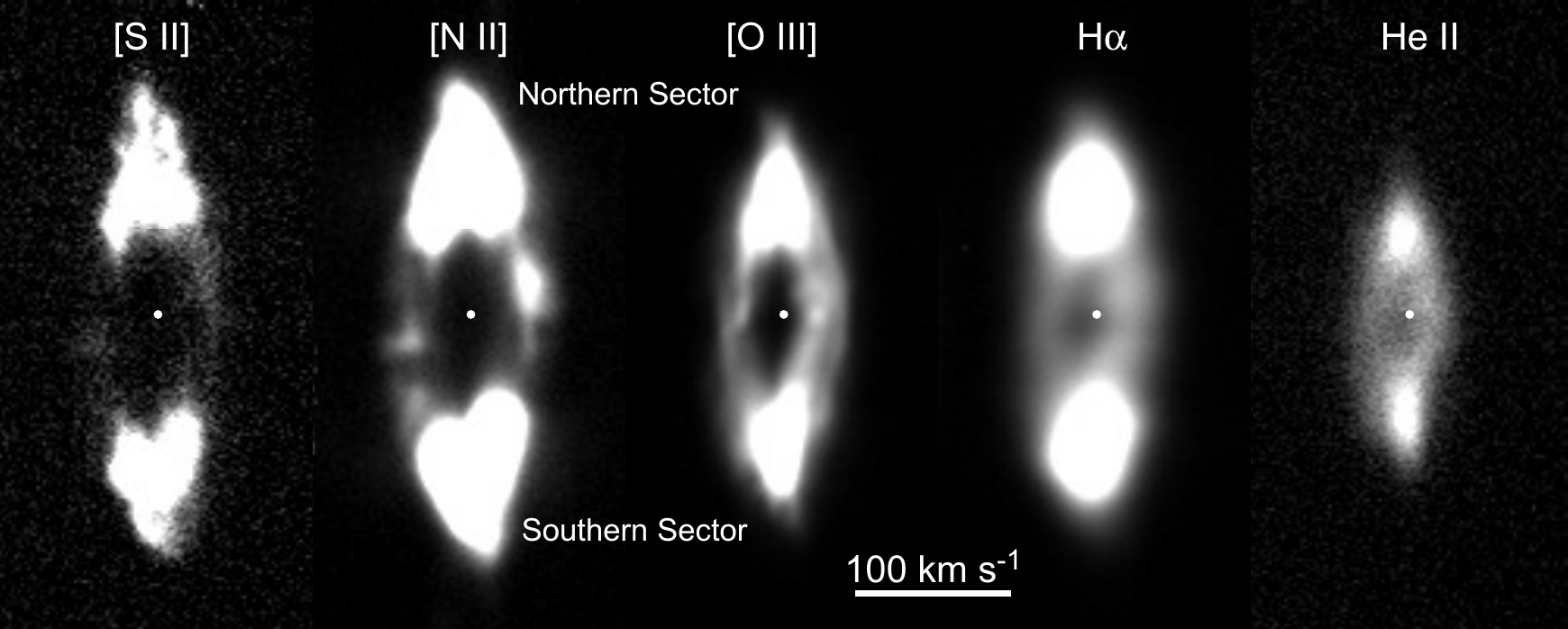}
\caption{Averaged spectra over the entire nebula are shown in the same style as Figure~\ref{fig:f12}, with the level of illumination set to show the features in the central region and all of the major emission-lines depicted. 
\label{fig:f13}}
\end{figure}

In Figure~\ref{fig:f13} we see that there are well defined small structures in the \nii\ spectrum near the central star and having velocities of -38 \kms (southern sector) and +38 \kms\ (northern sector). We identify these with the inner region features discussed in \S\ \ref{sec:inner} that are broad streaks in He~II and small arcs in \nii\ and \htwo. 
The \nii\ velocities are close to \Vexp\ = 34 \kms\ identified above for the Lobes. We see He~II emission at -25 \kms\ associated with the joint features Ea+Eb. The inner features projected near the Central Star are probably associated with the Lobes, rather than being portions of the Inner Halo seen in projection. These features are certainly optically thick to LyC photons as they have He~II emission, low ionization, and corresponding \htwo\ emission. The difference in radial velocities fit into the usual pattern of increasing spatial velocity with increasing distance from the central star and further associate these inner features with the Lobes. 
It is likely that the inner features represent Inner Halo gas being overtaken and accelerated by the more rapidly expanding Lobes.

\subsection{A Comparison of  our model with previous studies.}
\label{sec:CompPrevious3D}

As one of the classic planetary nebulae \ngc\ has been the subject of multiple attempts to build a 3-D model and it is important to compare the model presented in \S\ \ref{sec:New3D} and \S\ \ref{sec:Halo} with them. Our new model is essentially an elaboration of the work based on modeling through an assumption of a linear expansion relation applied to flux calibrated high velocity resolution data (OSH) and reaches similar conclusions on the structure of the Main Ring. A similar model to our new model was presented in the study of Steffen et al. (2007), who used the results from a few long-slit spectra, creating from them a 3-D model using the SHAPE modeling program, but publishing the results in a short conference proceeding. Our larger amount of data (many more spectra, multiple states of ionization, \htwo\ images, high resolution HST imaging) has allowed us to refine and expand the model.

One can trace the evolution of the 3-D model for \ngc\ from the recognition that there was an equatorial concentration of material producing an ionization bounded object \citep{mas90}, addressing the Lobe structure \citep{bry94}, and application of high resolution spectroscopy GMC, OSH, \citep{ste07}. It is difficult to place into this sequence the study of Kwok et al. (2008) based entirely on imaging and drawing on the likely similarity of form but difference of orientation of NGC~6853.

\subsection{Comparison with models of the Helix Nebula}
\label{sec:Helix}

There is a striking resemblance of \ngc\ and the Helix Nebula (NGC~7293). As noted in \S\ \ref{sec:Knots}, the factor of about three greater distance to the \ngc\  means that we don't see the fine-scale features as clearly. However, the large-scale appearance of both nebulae has been clearly defined in a wide variety of ions \citep{ode98,omm04}. In the most complete analysis of the 3-D structure of the Helix Nebula, O'Dell et al. (2004) concluded that the apparent incomplete double-ring structure and multiple apparent ionization fronts that are seen there (also in the \ngc) were due to two rings of material tilted almost perpendicular to one another, i.e. that the nebula was quadrapolar, rather than the common bipolar structure frequently seen. Meaburn et al. (2005) used the available velocities as interpreted by the XShape code and derived a bipolar model of the Helix Nebula where much of the radiation arises from seeing the Lobes in projection along the line of sight. The true 3-D model for the Helix Nebula is uncertain because of the peculiar and incompletely characterized velocity pattern.  Because of the much lower surface brightness of the Helix Nebula we do not have a complete velocity map of the Helix Nebula in emission-lines from a wide range of ionization states, so that there is a greater uncertainty in the true Helix Nebula model than in the \ngc\ model and the Helix Nebula 3-D model may change as a full data set becomes available.

\section{Summary and Conclusions}

We have made an in-orbit calibration of the specialized (targeting the study of gaseous nebulae) narrow-band filters incorporated into the WFC3 of the HST using ground-based spectra of  \ngc\ as a reference source. The flux-calibrated images of \ngc\ were then used with ground-based high velocity resolution spectra to determine  a more realistic 3-D model of it. Our major conclusions are listed below.

1. The in-orbit calibration of the narrow-band WFC3 filters are in good general agreement with that expected from the pre-launch determination of the filter properties  with the exception of a small systematic difference probably due to the standard stars used, differences arising from emission-lines that fall onto the blue-profile of the filters (where the transmission profile differs most), and a more deviant and unexplained difference in the F469N filter measuring the He~II 468.6 nm line.

2. Calibration constants are presented that should allow other observers to make flux-calibrated WFC3 emission-line images that include both corrections for contaminating extraneous lines (where present) and the underlying continuum.

3. Radial profiles through the Main Ring indicate the presence of multiple ionization fronts. These features in the WFC3 images, when combined with radial velocity information, argue that the Main Ring is a highly irregular ionization-bounded disk lying almost in the plane of the sky. This equatorial disk is not circular, which explains why the nebula is brightest in directions of the minor axes. 

4. The Main Ring's inner region marks the presence of a pair of extended Lobes pointed almost at the observer. These Lobes are radiation-bounded and have higher expansion velocities than the Main Ring disk.

5. The previously known faint halo features are almost certainly fossil radiation, i.e. this is gas that was photoionized during an earlier evolutionary state, when the central star was sufficiently luminous in ionizing photons that the entire equatorial disk was ionized. 

6. There are numerous knots and these appear to be located near the current ionization boundaries. Unlike the Helix Nebula, many of these knots are not highly symmetric. This may indicate that they have not been subject to the central star's radiation field long enough for shaping.

 \acknowledgments
We are grateful to David Thompson of the Large Binocular Telescope Observatory for providing copies of his unpublished LBTO \htwo\ data taken with the LUCI1 instrument and
to Antonio Peimbert for several fruitful discussions.

 GJF acknowledges support by NSF (0908877; 1108928; and 1109061), NASA (10-ATP10-0053, 10-ADAP10-0073, and NNX12AH73G), JPL (RSA No 1430426), and STScI (HST-AR-12125.01, GO-12560, and HST-GO-12309).  MP received partial support from CONACyT grant 129553. WJH acknowledges financial support from DGAPA--UNAM through project PAPIIT IN102012. CRO's participation was supported in part by HST program GO 12309. 

{\it Facilities:} \facility{HST {(WFC3)}}

\appendix
\section{Calibration of the Emission-Line Filters}
\label{AppendixA}

Calibration of the emission-line filters means the establishing of a protocol that allows converting signals  (\rfilter, electrons pixel$^{-1}$ \pers) from extended sources imaged with the WFC3 into corresponding monochromatic surface brightnesses (S, \sbunits). This has been done using a bright extended emission-line source, NGC~6720, as a reference object. Regions within that object have been subjected to ground-based spectroscopy that has derived the monochromatic surface brightness through reference to standard calibration stars. The product of the WFC3 calibration are primary constants that allow conversion of \rfilter\ into surface brightness units and secondary constants that allow correction for non-targeted emission-lines that affect the signal for a given filter and for the underlying continuum. As we will establish, the contamination of the signals can lead errors in the derived surface brightness of more than a factor of two. 

In the absence of contaminating signal, the relation between the count rate (\rfilter) and the surface brightness of the extended source will be 

\begin{equation}
\rm S = R_{filter} / K_{filter} 
\end{equation}

\Kfilter\ = A~x~T~x~$\Omega$~where A is the light gathering area of the HST, T again is the throughput of the telescope plus camera, and $\Omega$ is the solid angle of one WFC3-UVIS pixel. Throughput is the product of the transmission of the optical train (including the telescope and instrument) multiplied by the quantum efficiency of the detector. Adopting an 86\%\ illumination for the 2.4-m HST and a pixel size of 0.04\arcsec\ x 0.04\arcsec\ (appropriate for the pipeline-processed images that were our starting material), one would expect \Kfilter\ = 14.632 x 10$\rm ^{-10}$ x T, where T is the throughput of the filter for the targeted emission line.  in \S\ \ref{sec:cont} we present the results of determination of \Kfilter\ and the contamination terms for each filter. 

\subsection{Ground-based Spectra Used as Reference Signals}
\label{sec:SPM}

The ground-based spectra we used are described in O'Dell et al. (2009) and O'Dell (2009). A 3.93\arcsec\ wide long slit was placed across the nebula at position angles (PA) of 60\arcdeg\ and 150\arcdeg\ as shown in Figure~\ref{fig:f2} . The astronomical seeing was 2.3\arcsec\ FWHM  (Full Width at Half Maximum) during the observations and the resolution of the spectra was 0.94~nm. Through near-simultaneous observations of the flux standard star BD+25~3941, IRAF tasks were used to extract emission-line surface brightnesses for the lines in and near the bandpasses of the WFC3 filters. Five samples A-J (also shown in Figure~\ref{fig:f2}) were taken in the regions of the spectra that overlapped all of the WFC3 images.

\subsection{Matching the Imaging and Spectroscopic Samples}
\label{sec:matching}

The portions of the WFC3 image closely matching the spectral samples were identified, with the irregular
length of the samples reflecting the scaling of pixel sizes (1.05\arcsec\ for the spectra, 0.04\arcsec\ for the pipeline-processed WFC3 images).  In order to match the images, the HST images were blurred using the IRAF task "gauss" to produce the same 2.3\arcsec\ FWHM as the astronomical seeing that prevailed during the spectroscopic observations. The average count rate \rfilter\ was then determined. 

\subsection{Correction for Continuum and Contaminating Emission-lines}
\label{sec:cont}

The simple application of Equation 1 can result in significant photometric errors unless a correction has been made for signal arising from contaminating lines and the underlying continuum. The observed \rfilter\ value must in general be divided by a correction term \rcorrfilter\ defined as 


\begin{equation}
\rm r_{filter} = 1 + \frac{W(filter)}{EW} + \Sigma \frac{T(contaminating~line)}{T(targeted~line)} \frac{S(contaminating~line)}{S(targeted~line}
\end{equation}

with the first fraction being the correction term for the underlying continuum and the second being the correction term for non-targeted emission-lines falling within the bandpass of the filter.
Without these corrections one cannot in general determine accurate photometry of emission-line objects even with the narrow-band filters. For purposes of the filter calibration we have determined the correction terms using the ground-based spectra. For the F502N and F658N images \rcorrfilter = 1.00, but for other filters, \rcorrfilter\ varied from a few percent over unity to as large as 2.84.  

There will always be a component of the signal in a WFC3 narrow bandpass due to the underlying nebular and scattered starlight continuum. Its strength relative to the signal from the targeted emission line will be W(filter)/EW, where W is the bandpass rectangular width (the equivalent width divided by the maximum throughput for that filter) and EW is the equivalent width of the targeted line (the surface brightness of the line divided by the surfaced brightness in the underlying continuum). We adopted the pre-launch determined values of W from Table 6.2 of the WFC3 Instrument Handbook, except increasing the values slightly because of the
broadening due to short wavelength shift of the filter profile. The amount of the shift was guided by the determination from emission-lines falling on the short wavelength portion of the filter profile described below.  The adopted values of W were FQ436N (4.5~nm), FQ437N (3.3~nm), F469N (5.31~nm), F487N (6.2~nm), FQ575N (2.06~nm), FQ672N (2.21~nm), F673N (12.31~nm), and FQ674N (2.11~nm). It was not necessary to adopt a value of W for F502N, F656N, and F658N because the observed continuum was so weak that the continuum correction was less than 0.01). 

Five of the filters were affected by contaminating emission-lines and their T's are shown in Figure~\ref{fig:f2}. In the case of this type of contamination, the fraction of the signal from the targeted line will be [T(contaminating line)/T(targeted line)]~x~[S(contaminating line)/S(targeted line)]. For the F469N filter the contaminating line was the 471.1~nm line of [Ar~IV]. Since the ionization energy necessary to produce Ar$^{+++}$ is higher than that producing He$^{++}$, the ratio is expected to be rather constant across the nebula (which was the case), The throughput ratio was taken to be 0.70 and is not affected by the uncertainty of the shape of the short wavelength side of the filter profile. This is not the case for the F656N and FQ674N filters and the throughput ratios had to be determined using relevant \rfilter\ values. In both cases we determined T(contaminating line)/T(targeted line) by assuming that the continuum corrected signal from the targeted line would be proportional to the T value at that line. We have ignored the presence of the 656.0 nm line of He$^{++}$ within the F656N filter since the emissivities of that line \cite{ost06} indicate that in the Main Ring it should be less than 1 \%\ the strength of \Ha. The contaminating line for FQ674N (\sii~ 6716) was the targeted line of an adjacent filter (FQ672N). We then found the solution for FQ674N [T(contaminating)/T(targeted) ] to be 0.07, indicating that there is not a m/peasurable  blueward shift of the short-wavelength profile for FQ674N. A similar analysis was done of the contamination of the F656N filter by the \nii\ 654.8~nm line, except we drew on the F658N images (targeting the 658.3~nm  member of the \nii\ doublet and utilized the fact that the S(6548)/S(6583) ratio is intrinsically constant and was determined to be 0.331$\pm$0.008 in our set of spectra. The derived ratio of T(contaminating)/T(targeted) was 0.15 for the F656N filter, corresponding to a profile shift of about 0.14~nm. 

The FQ437N filter is primarily contaminated by the usually much weaker HeI 438.8~nm emission line. Fortunately, this line should arise from the same ionization stratification zone as the \oiii\ emission and it should be relatively constant with respect to \oiii\ 436.3~nm. Using the published throughputs,  we found T(contaminating line)/T(targeted line) = 0.10 and from the spectroscopically determined line ratios the correction factor could be determined.

The FQ436N filter is affected by both the weak 438.8~nm line, in this case T(contaminating line)/T(targeted line) = 0.75,
and the strong \Hg\ 434.0~nm line. Again it was assumed that the corrected \rfilter\ ratios (for FQ436N and FQ437N) should be the same as their relative throughput values and it was determined that T(contaminating line)/T(targeted line) for \Hg\ was 0.33$\pm$0.05. This corresponds to a blueward shift of the FQ436N short wavelength profile of 0.28~nm.  

\subsection{Derivation of the Correction Terms for NGC~6720}
\label{sec:general}

In order to determine emission-line surface brightnesses for parts of the nebula that were imaged but not subject to spectroscopy, we have determined the relationship between the correction terms and relevant observed parameters. For example, the EW of a targeted line will be expected to correlate with R$_{filter}$/R$_{547}$, where R$_{547}$ is the signal from the F547M filter, which is dominated by continuum emission. Such an assumption assumes that the spectral distribution of the continuum is relatively constant (the spectra indicate that this is the case) and that variations due to reddening are minor. Similarly, \rnii / \rha\ will be a guide to the contamination of the F656N filter by the 654.8~nm \nii\  line.
A least squares fit was made to the correction terms and the results for the predicted correction term (r$\rm_{pred}$) are also given in Table 1. A measure of the uncertainty of this approach is indicated by the root mean squared values of the difference in r$\rm_{obs}$ (the value of $\rm r_{filter}$ obtained from the spectroscopy) and r$\rm _{pred}$. These varied from 0.016 to 0.045 for the filters useful for the present temperature fluctuations study, but were larger (0.077) for the problematic F469N filter and 0.381 for the FQ436N filter.
The He~II F469N filter was not necessary for our temperature fluctuations study and we conclude that the very large  contamination of the FQ436N by the \Hg\ line render it unusable for obtaining accurate information on the \oiii\ 436.3 nm line. 

These correction terms ($r\rm_{fit}$) should generally be useful for the study of nebulae similar to the Ring Nebula, where the key factors are high ionization and low densities. The F502N, F656N, and F658N are relatively insensitive to the contamination terms. However, detailed application of this general approach for the other filters in other objects requires confirming the contamination terms. The more general procedure necessary for calibration the contamination terms in objects not similar to \ngc\ is described in \S\ \ref{sec:rmpred}.

\subsection{Derived \Kfilter\ Values and Comparison with Expectations}
\label{sec:Kfilter}

The \Kfilter\ values derived from the ten sampled regions common to the San Pedro Martir spectra and the WFC3 images are presented in Table~4. A comparison with the values predicted (K$\rm_{pred}$) from the assumptions of the throughputs in the WFC3 Instrument Handbook, a pixel size of 0.04\arcsec\ x 0.04\arcsec\ and 86\%\ illumination of the HST's primary mirror are also given. Exuding the results for F469N, the average \Kfilter/K$\rm_{pred}$ = 1.14$\pm$0.09, with the deviations expected from the uncertainties in \Kfilter\ for individual filters being 0.08. The Instrument Handbook throughput of the wide-band WFC3 filters were calculated from in-orbit observations of standard stars and the throughputs for the narrow-band filters were scaled from these results on the basis of the ground-based measures of both. The small systematic difference in \Kfilter/K$\rm_{pred}$ is likely to be due to a slightly different calibration standard for the star BD+25~3941 and the stars used by the STScI in their on-orbit calibration.

The greatest deviation in \Kfilter/K$\rm_{pred}$ is for the F469N filter, where it is 1.71$\pm$0.14. The deviation is remarkable as it argues that the filter has increased in throughput between the time of the ground-based calibration and the in-orbit determination of \Kfilter! There are significant variations in the correction term for this filter, with a range from r = 1.164 to r = 2.035, but there is no systematic difference in the value of \Kfilter\ with the value of r. This makes it unlikely that the deviation is caused by systematic uncertainties in the variation with the correction factor r. The filter is reported to operate in the first order and the discrepancy could indicate an ultraviolet leak near 234.5 nm. The strongest line in this region in planetary nebulae is the He~II 238.1 nm line; but, it falls outside of the wavelength range that is one-half of the nominal F469N filter's bandpass. If the F469N filter is actually operating in the second order, then its third order bandpass would be almost exactly centered on the intrinsically strong 313.2 nm O~III line than is pumped by the He~II Ly$\alpha$ resonance line \citep{ost06} and this could account for the unexpectedly large signal and derived \Kfilter.

The spread in the determination of \Kfilter is much larger than expected from photon statistics since the minimum total signal was 66400 electrons (sample F, FQ674NN). A similar situation was found when the same basic approach was applied to calibration of the HST WFPC2 \citep{ode09} and the HST ACS \citep{ode04} instruments. The reason is probably the failure of a gaussian blurring of the HST images to adequately match the conditions that applied during the spectroscopic observations.

\subsection{The General Filter Calibration Values for  r$\rm_{pred}$ Expected from the Filter Characteristics}
\label{sec:rmpred}

One can derive the behavior of r$\rm_{pred}$ from the instrument handbook values of the throughput together with the corrections for the short wavelength broadening of the filters (mostly affecting the contamination of F656N by the \nii\ 654.8 nm line and FQ674N by the \sii\ 671.6 nm line).  These values should be valid for a variety of nebulae and physical conditions. In the absence of supporting spectra that allow empirical determination of the correction terms, these values of r$\rm_{pred}$ should be useful for studies of other nebulae.

\subsubsection{The Simplest Filters}
\label{sec:simplest}

In the case of the filters having only contamination by the underlying continuum, r$\rm_{pred}$ takes its simplest form 

\begin{equation}
\rm r_{pred} = 1 + (\frac{1}{\ratio}\frac{\Wc}{\W}\frac{\Tc}{\Tm}\frac{R_{filter}}{R_{F547M}} - 1)^{-1},
\end{equation}

where W$\rm _{F547M}$ and W$\rm_{filter}$ are the bandpass rectangular widths of the F547M and the line filters, \Tc\ and \Tm\ are the maximum throughputs of the F547M and the line filters, and R$\rm _{F547M}$ and $\rm R_{filter}$ are the count rates in the F547M and the line filters, while k is the flux ratio of the continuum (in wavelength intervals) at the line filter wavelength and at 547 nm.  

In the case of the F487N filter, r$\rm_{pred}$ = 1 + [10.91  k\per\ (\rhb / \rcont)   - 1]\per. 

For the F502N filter, r$\rm_{pred}$ = 1 + [10.31 k\per\ (\roiii / \rcont)   - 1]\per. 

For the F658N filter, r$\rm_{pred}$ = 1 + [22.41  k\per\ (\rnii / \rcont)   - 1]\per. 

For the FQ672N filter, $\rm_{pred}$ = 1 + [35.58  k\per\ (\rslow / \rcont)   - 1]\per.

For the F673N filter, r$\rm_{pred}$ = 1 + [5.73  k\per\ (\rsii / \rcont)   - 1]\per.

When using our spectra of NGC~6720 to test these relations we find r.m.s. deviations of 1\%\ for F502N and F658N, 2 \%\ for F487N and FQ672, and 5 \%\ for F673N.  

\subsubsection{The More Complex Filters}
\label{sec:complex}

The situation is more complex for those filters where there is a contaminating line producing a significant addition to the targeted line. The problem is tractable when there is a direct observational measure of the strength of the contaminating line. In the case of the F656N  \Ha\ filter, which has a contribution from the \nii\ 654.8 nm line, one can use the signal from the F658N filter, which targets the other member of the \nii\ doublet at 658.3 nm. Similarly, one can correct for the \sii\ 671.6 nm line that contaminates the FQ674N signal from observations in the FQ672N filter, which targets the 671.6 nm line. The resulting equation for the F656N r$\rm_{pred}$  is 

\begin{equation}
\rm r_{pred} = 1 + (37.12~k ^{-1} \frac{R_{F656N}}{R_{F547M}} - 1)^{-1} + 0.04488~\frac{R_{F658N}}{R_{F656N}},
\end{equation}

where the first term is the correction for the underlying continuum and the second for the contaminating line, under the assumption that the continuum contamination of the F658N filter is negligibly small, which is valid in this application. For the FQ674N filter one must also include the continuum correction term to the reference FQ672N signal and the equation becomes

\begin{equation}
\rm r_{pred} = 1 + (49.42~k ^{-1} \frac{R_{FQ674N}}{R_{F547M}} - 1)^{-1} + 0.05467~\frac{R_{FQ672N}}{R_{FQ674N}}\frac{1}{1 + (35.58~k^{-1} R_{FQ672N}/R_{F547M} - 1)^{-1} },
\end{equation}

where again the first term is the continuum correction and the second term is the contamination line correction but with the continuum correction for the FQ674N filter included. Using the NGC~6720 spectra for reference gives an r.m.s. deviation for F656N of 1~\%\ and 3~\%\ for F674N.

In the case of the FQ575N filter there is the need only for the continuum correction term, as in Equation A1. However, for this filter the
continuum term is much larger than for the five continuum-only corrected filters cited above and the predictions for r$\rm_{pred}$ are correspondingly more sensitive to the assumed values of the throughput and the color of the continuum.  Adopting k~=~1.0 and the nominal filter characteristics, one finds for the lower ionization samples (B, C, D, E, H, I, J) r$\rm_{pred}$/r$\rm_{obs}$ = 1.18$\pm$0.10.  By experimenting with the scale factor in the contamination term, we found that increasing this by a factor of 1.4 gave r$\rm_{pred}$/r$\rm_{obs}$ = 1.00$\pm$0.04. In this case the general expression becomes

r$\rm_{pred}$ = 1 + [50.44  k\per\ (\rniiup / \rcont)   - 1]\per.

The reason for the need for this scale factor is uncertain, but well defined.  The average k for these seven samples was 0.91$\pm$0.05, so that employing the observed color of the continuum would only increase the discrepancy. In the light of the relatively small dispersion after the correction is made, it is advised that the above relation be used in the study of other nebulae. The similarly small deviation in the comparison of $\rm  r_{obs}-r_{fit}$ (r.m.s.~=~0.035) indicates that we have an accurate guide for study of NGC~6720.  

In the case of the FQ437N filter there is the need for both a continuum and contaminating line correction term. Unfortunately, there is not a direct measurement relevant for applying the contaminating line correction and one must rely on spectra for a guide. For the higher ionization NGC~6720  samples (A, B, C, D, F, G, H, I), the ratio of the 438.8 nm and 436.3 nm lines is 0.10$\pm$0.03. Since the ratio of transmissions is only 0.10, this necessitates a correction term of 1~\%. Similar to the situation for FQ575N, we find poor agreement (r$\rm_{pred}$/r$\rm_{obs}$ = 1.23$\pm$0.11) for the nominal filter characteristics, but when a scale factor of 1.45 is applied, this becomes r$\rm_{pred}$/r$\rm_{obs}$ = 1.01$\pm$0.03.  Adopting this correction term, we have for the FQ437N filter

r$\rm_{pred}$ = 1.01\{1 + [36.79  k\per\ (\roup / \rcont)   - 1]\per \}.

We expect the line correction to be dependent on the ratio of the 438.8 nm and 436.3 nm flux. Although well defined and small in the case of the high ionization nebula NGC~6720, the ratio is much larger for a low ionization object like the the Orion Nebula (NGC~1976), where the ratio is about 0.45 \citep{bal00,est04}, in which case the above multiplier would be 1.045. Fortunately, our empirical relation for NGC~6720 given in Table 1 gives relatively small errors, $\rm  r_{obs}-r_{fit}$ (r.m.s.~=~0.044)

The F469N filter is affected by both a continuum contamination and 471.1 nm emission of [Ar~IV], the latter occurring at a throughput of 0.70 \%\ that of the primary 468.6 nm line.  For determination of r$\rm_{pred}$ we have studied the six innermost samples (A-C, and F-H) where He~II emission is the strongest. Within these samples the flux ratio of contaminating (471.1 nm) to targeted (468.6 nm) lines was nearly constant at 0.083$\pm$0.007.  and the ratio of observed to predicted correction factors was near unity with only a 1.1X increase in the continuum correction term, which led to the relation 

 r$\rm_{pred}$ = 1 + 0.058+ 0.07347 (\rcont / \rHeII)
 
 with an r.m.s. dispersion of 0.017. In the case of a lower ionization nebula both the targeted and contaminating lines will be intrinsically much weaker. High resolution studies of the Orion Nebula indicate that He~II line is absent and the [Ar~IV] line at 471.1 nm is only 0.062 \%\ \citep{bal00} and 0.096 \%\ \citep{est04} of \Hb. In similar low ionization objects the F469N filter arguably becomes a measure of the continuum. As discussed previously in a study using the similar WFPC2 F469N filter \citep{ode03} in low ionization objects contamination of the F469N filter can occur from [Fe~III] at 470.2 nm and He~I at 471.3 nm. Adopting the average of the high resolution studies, the [Fe~III] line flux is 0.19 \%\ of \Hb\ and the He~I line flux is 0.71 \%\ of \Hb. The equivalent width of the \Hb\ emission-line is about 50 nm in the central Orion Nebula (owing to a strong component of scattered starlight) \citep{oh11}, which means that the 470.2 nm and 471.3 nm lines would contribute about 10 \%\ of the F469N continuum signal, making it difficult to use the F469N filter as a continuum reference even in low ionization nebulae unless there is an object-specific spectrum-determined line correction.

The arbitrary nature of the scale-factors for the continuum corrections in the diagnostically important FQ437N and FQ575N filters and the F469N filter  are additional arguments that empirical determination of them from spectra of the program object is strongly recommended, rather than blindly using the r$\rm_{pred}$ values. The magnitude of the continuum contributions to the correction factor r$\rm _{filter}$ are larger for the former two filters (average 0.40 with a range of 0.21-0.99), whereas for the F469N filter they are smaller (average of 0.19 with a range of 0.10-0.242). The deviation is largest where the calculation is most dependent upon an accurate value of W$\rm _{filter}T_{F547M}$/W$\rm _{F547M}T_{filter}$. This suggests that these ratios deviate from those inferred from the pre-launch determined values given in the WFC3 Instrument Handbook.



\begin{deluxetable}{llllc}
\tabletypesize{\scriptsize}
\tablecaption{The Derived Calibration Coefficients* \Kfilter, K$_{pred}$,  r$\rm_{fit}$ and Residuals}
\tablewidth{0pt}
\tablehead{
\colhead{Filter} &
\colhead{\Kfilter} &
\colhead{\Kfilter /K$\rm_{pred}$} &
\colhead{r$\rm_{fit}$} &
\colhead{r.m.s. ($\rm  r_{obs}-r_{fit}$)}}
\startdata
FQ436N & 3.42$\pm$0.27 & 1.23 & ---- & ---- \\  
FQ437N & 3.75$\pm$0.47 & 1.30 & 0.983+0.040541xR$_{547}$/R$_{437}$ & 0.044 \\
F469N    & 4.61$\pm$0.40 & 1.71 & 0.888+0.136908xR$_{547}$/R$_{469}$$^{\#}$ & 0.085 \\
F487N    & 4.08$\pm$0.31 & 1.13 & 1.070 & 0.005 \\
F502N    & 3.59$\pm$0.17 & 1.04 & 1.00 & ---- \\
FQ575N & 4.28$\pm$0.40 &  1.28 & 0.970+0.029512xR$_{547}$/R$_{575}$ $^{\#\#}$& 0.035 \\
F656N    & 3.79$\pm$0.16 &  1.10 & 1.003+0.046488xR$_{658}$/R$_{656}$& 0.003 \\
F658N    & 3.95$\pm$0.40 &  1.05 & 1.00 & ----- \\
FQ672N & 3.86$\pm$0.27 & 1.09  & 1.038+0.015220xR$_{547}$/R$_{672}$ & 0.035 \\
F673N    & 3.85$\pm$0.24 & 1.14 & 0.929+0.226527xR$_{547}$/R$_{673}$ & 0.045 \\
F674N    & 3.00$\pm$0.25 & 1.08 & 1.074+0.023674xR$_{547}$/R$_{674}$ & 0.016 \\
\enddata
\tablecomments{~*All \Kfilter\ values are in units of 10$^{-10}$ electrons cm$^{2}$ ster photons$^{-1}$, while r$\rm_{fit}$ and r$_{obs}$\ are dimensionless.
$^{\#}$Only samples with R$_{547}$/R$_{469}$$\leq$8 and R$_{502}$/R$_{487}$$\geq$9 were used in this solution.
$^{\#\#}$Only samples with R$_{547}$/R$_{575}$$\leq$20 were used in this solution.}
\end{deluxetable}

\end{document}